\documentclass[12pt,a4paper]{article}   
\usepackage{epsf}

\newlength{\HFPP}       \HFPP5.4mm
\makeatletter
\@addtoreset{equation}{section}
\makeatother

\begin{document}
\begin{titlepage}
\def\thefootnote{\fnsymbol{footnote}}

\begin{flushleft}
\setlength{\baselineskip}{13pt}
OUTP-95-67S \hfill December 1995 \\
ITP-UH-28/95 \hfill revised July 1996 \\
ITP-SB-95-48 \\
YITP-96-26
\end{flushleft}

\vspace*{\fill}
\begin{center}
{\Large Determinant representation for a quantum correlation function\\[10pt]
        of the lattice sine--Gordon model} \\
\vfill
\vspace{1.5 em}
{\sc Fabian H.\,L.\,E\ss{}ler}\footnote{%
e-mail: {\tt fab@thphys.ox.ac.uk}}\\
{\sl Department of Physics, Theoretical Physics\\
     1 Keble Road, Oxford OX1 3NP, Great Britain}\\[8pt]
{\sc Holger Frahm}\footnote{%
e-mail: {\tt frahm@itp.uni-hannover.de}}\\
{\sl Institut f\"ur Theoretische Physik, Universit\"at Hannover\\
     D-30167~Hannover, Germany}\\[8pt]
{\sc Alexander R.\ Its}\footnote{%
e-mail: {\tt itsa@math.iupui.edu}}\\
{\sl Department of Mathematical Sciences,\\
        Indiana University-Purdue University at Indianapolis (IUPUI),\\
        Indianapolis, IN 46202--3216, U. S. A.}\\[8pt]
{\sc Vladimir E.\ Korepin}\footnote{%
e-mail: {\tt korepin@insti.physics.sunysb.edu}}\\
{\sl Institute for Theoretical Physics,
     State University of New York at Stony Brook\\
     Stony Brook, NY 11794--3840, U. S. A.\\ and \\
Yukawa Institute for Theoretical Physics, Kyoto
University, Kyoto 606, Japan\\
and\\
 Sankt Petersburg Department of Mathematical Institute of
Academy of Sciences of Russia}\\

\vfill
ABSTRACT
\end{center}
\begin{quote}
We consider a completely integrable lattice regularization of the
sine--Gordon model with discrete space and continuous time. We derive a
determinant representation for a correlation function which in the
continuum limit turns into the correlation function of local fields. The
determinant is then embedded into a system of integrable
integro--differential equations. The leading asymptotic behaviour of
the correlation function is described in terms of the solution of a
Riemann Hilbert Problem (RHP) related to the system of
integro--differential equations. The leading term in the asymptotical
decomposition of the solution of the RHP is obtained.
\end{quote}
\vfill
\setcounter{footnote}{0}
\end{titlepage}

\def\sG{sine--Gordon}
\def\a{\alpha}
\def\b{\beta}
\def\g{\gamma}
\def\k{\kappa}
\def\dv{|0)}
\def\ddv{(0|}
\def\nddv{{(\tilde 0|}}
\def\La{\Lambda}
\def\be{\begin{equation}}
\def\ee{\end{equation}}
\def\bea{\begin{eqnarray}}
\def\eea{\end{eqnarray}}
\def\r#1{(\ref{#1})}
\def\vac{|0\rangle}
\def\nn{\nonumber\\}
\def\l{\lambda}
\def\lt{{\widetilde{\lambda}}}
\def\m{\mu}
\def\emph#1{{\sl #1}}
\def\textbf#1{{\bf #1}}
\def\A{{\exp(-2\La)}}
\def\B{{\exp(2\La)}}
\def\wh#1{\widehat{#1}}
\def\wti#1{\widetilde{#1}}
\def\Op{{\cal O}}
\def\I{\left(\matrix{1&0\cr 0&1\cr}\right)}
\def\tr{\mbox{tr}}
\def\dual{\stackrel{\circ}{\varphi}}
\def\fl{(\lambda)}


\section{Introduction}
The \sG\ model is completely integrable (exactly solvable) both on the
classical and on the quantum level \cite{akns:73}%
\nocite{tafa:75,zatf:75,kane:78,tafa:79,fata:book}--\cite{stf:79}. We shall
write the \sG\ equation in the following form:
\begin{equation}
   {\partial^2 \over \partial t^2} u(x,t) -
   {\partial^2 \over \partial x^2} u(x,t) +
   {m^2\over \beta} \sin \beta u(x,t) = 0\ .
\label{sg:eq}
\end{equation}
Here $m$ is a mass, $\beta$ is the coupling constant. For later convenience
we also introduce 
\[
   \gamma = {\beta^2\over 8}\ .
\]
In the classical case $u(x,t)$ is an function of two variables, $x$ and $t$
are space and time coordinates. In the quantum case $u(x,t)$ is a local
quantum field of the \sG\ model. The Hamiltonian reads
\begin{equation}
    {\cal H} = \int dx \left( {1\over2} p^2 + {1\over2}(\partial_x u)^2
	+{m^2\over\beta^2} (1-\cos\ \beta u) \right)\ .
\label{sg:hamil}
\end{equation}
Momentum and topological charge are given by
\begin{equation}
   P = - \int dx\ p\ \partial_x u\ , \qquad 
   Q = {\beta\over2\pi} \int dx\ \partial_x u\ .
\label{defcon}
\end{equation}
Here $p(x,t) = \partial_t u(x,t)$ and $u(x,t)$ satisfy Poisson brackets
$\left\{ p(x) , u(y) \right\} = \delta(x-y)$. Equation (\ref{sg:eq}) has a
Lax representation and a classical $r$-matrix
\cite{akns:73}\nocite{tafa:75,zatf:75,kane:78,tafa:79}--\cite{fata:book}.
After quantization, the fields $u$ and $p$ satisfy canonical commutation
relations $\left[ u(x) , p(y) \right] = i\ \delta(x-y)$.  The physical
ground state $|\Omega\rangle$ of the quantum system can be obtained by
filling the Dirac sea of negative energy pseudoparticles \cite{stf:79}.

Let us now consider the quantum operator
\begin{equation}
   \exp\left( \alpha Q(x) \right) 
	= \exp\left\{ {\alpha\beta\over2\pi}\bigl(u(x)-u(0)\bigr)\right\}\ ,
   \qquad
   Q(x) = {\beta\over2\pi} \int_0^x dz \partial_z u(z)\ ,
\label{expaq}
\end{equation}
where $Q(x)$ measures the topological charge on the interval $[0,x]$.
In this paper we show how to represent the correlation function
\begin{equation}
   \langle \Omega | \exp\left( \alpha Q(x) \right) | \Omega \rangle
\label{corrf}
\end{equation}
as the determinant of an integral operator (in fact we shall see below,
that the coefficient $\alpha$ in (\ref{corrf}) needs to be renormalized).
Note that via differentiation with respect to $\alpha$ we can obtain
correlation functions of local quantum fields from \r{corrf}. We
shall consider the quantum version of (\ref{sg:hamil}) in the
region $\frac{\pi}{2}<\gamma<\frac{2\pi}{3}$ (many of our intermediate
results hold in larger regions of coupling constant). Note that
$\gamma\to0$ is the quasiclassical region of the \sG\ model and at
$\gamma=\pi/2$ the spectrum of the Hamiltonian is equivalent to free
fermions. 
To deal with the ultraviolet divergences of the continuum model we shall
employ a suitably chosen lattice regularization. 

The determinant representation then permits to describe the correlation
functions in terms of a system of integrable integro--differential
equations. These equations can be solved by means of a Riemann-Hilbert
problem which in turn enables one to obtain elementary formulas for the
asymptotics of the correlation functions. This program has first been
applied to the nonlinear Schr\"odinger equation in \cite{itsx:90} and
is described in detail in the book \cite{vladb} (see also
\cite{slav:89}). 

There has been previous work on determining correlation functions in
the \sG\ model. Form factors were determined by Smirnov in
\cite{smir:86,smirnov:92}. At the free fermionic point $\gamma=\pi/2$
a determinant representation of the correlation function (\ref{corrf})
has been constructed using the coordinate Bethe Ansatz in
\cite{itkt:92}. A description of a different correlator at the free
fermionic point through a Fredholm determinant (derived from a form
factor sum) which in turn satisfies an integrable differential
(sinh-Gordon) equation has been obtained in \cite{belc:94}. In this
paper we start the investigation of correlation functions in the \sG\
model for general $\gamma$, in particular away from the free fermionic
point in the framework of its solution \cite{izko:81,izko:82} by means
of the Quantum Inverse Scattering Method.

The plan of this paper is as follows: in section \ref{sec:lsg} we review
the integrable lattice regularization of the \sG\ model introduced in
\cite{izko:82}. The Algebraic Bethe Ansatz is formulated and the
construction of the ground state \cite{bogo:82} is discussed. In section
\ref{sec:qcf} we derive the determinant representation of the correlator
(\ref{corrf}) for the range of coupling constants stated above. As
this part of the analysis is very similar to the analogous problem for
the spin-$\frac{1}{2}$ Heisenberg XXZ model (which was treated in full
detail in \cite{efik:95}) we omit many details and only give an
account of the main steps without providing prrofs (which can be found
in \cite{efik:95}). In sections \ref{sec:cont} and \ref{sec:ide} we
embed the determinant representation into a system of integrable
integro-differential equations and in section \ref{sec:rhp} the
related Riemann-Hilbert problem is formulated and the leading
asymptotic behaviour of the correlation function is extracted.
\section{Lattice Sine--Gordon}
\label{sec:lsg}
\subsection{${\cal L}$-Operator}
We shall consider a lattice version of the \sG\ model which is also
completely integrable. It will have exactly the same $r$-matrix (both in
the classical and quantum case) as the continuous model. The elementary
${\cal L}$-operator of the LSG model is \cite{izko:81,izko:82}
\begin{equation}
  \label{lop}
  {\cal L}(n|\lambda) = \left(\begin{array}{cc}
  e^{-i{\beta p_n/ 8}} \rho_n e^{-i{\beta p_n/ 8}} &
     {1\over2}m\Delta \sinh(\lambda-i\beta u_n/2) \\[8pt]
  -{1\over2}m\Delta \sinh(\lambda+i\beta u_n/2) &
  e^{i{\beta p_n/ 8}} \rho_n e^{i{\beta p_n/ 8}}
  \end{array} \right)
\end{equation}
Here $\Delta$ ist the lattice constant and $p_n$, $u_n$ are the dynamical
variables on site $n$ of the lattice. In the quantum model they obey
canonical commutation relations $\left[ u_n, p_m\right] = i\delta_{nm}$.
Furthermore, we have introduced
\begin{equation}
  \label{defrho}
  \rho_n = \left( 1 + 2S\cos\beta u_n \right)^{1\over2}, \qquad
  S=\left({1\over4}m\Delta\right)^2\ .
\end{equation}
The symmetries of the ${\cal L}$-operator of the LSG model are expressed by
the identities (the asterisk means Hermitian conjugation of the quantum
operators)
\begin{equation}
\label{lsym}
   \sigma^y\ {\cal L}^*(n|\bar\lambda)\ \sigma^y = {\cal L}(n|\lambda)\ ,
   \qquad 
   \sigma^z\ {\cal L}(n|\lambda)\ \sigma^z = {\cal L}(n|\lambda+i\pi)\ .
\end{equation}
Its quantum determinant \cite{izko:81,izko:82} is
\begin{equation}
   \label{detqL}
   {\det}_q{\cal L}(n|\lambda) \equiv 1 + 2S\cosh 2\lambda\ .
\end{equation}

The ${\cal L}$-operator (\ref{lop}) satisfies the Yang-Baxter equation
\begin{equation}
  \label{ybe}
  R(\lambda,\mu) \left( {\cal L}(n|\lambda) \otimes {\cal L}(n|\mu) \right)
  = \left( {\cal L}(n|\mu) \otimes {\cal L}(n|\lambda) \right) 
       R(\lambda,\mu)\ .
\end{equation}
$R(\lambda,\mu)$ in Eq.\ (\ref{ybe}) is the standard \sG\ $R$-matrix given
by the following expression:
\begin{equation}
  \label{rmat}
  R(\lambda,\mu) = \left(\begin{array}{cccc}
   f(\mu,\lambda) & 0 & 0 & 0 \\
   0 & g(\mu,\lambda) & 1 & 0 \\
   0 & 1 & g(\mu,\lambda) & 0 \\
   0 & 0 & 0 & f(\mu,\lambda) \end{array}\right)\ .
\end{equation}
Here
\begin{equation}
  \label{def:fg}
  f(\mu,\lambda) = {\sinh(\mu-\lambda-i\gamma) \over \sinh(\mu-\lambda) }\ ,
  \qquad 
  g(\mu,\lambda) = -i {\sin\gamma \over \sinh(\mu-\lambda)}\ .
\end{equation}

In different sites of the lattice the matrix elements of ${\cal L}$
commute.  As usual in the Quantum Inverse Scattering method (QISM) we
define the monodromy matrix by taking products of the ${\cal L}$-operators
in matrix space:
\begin{equation}
   {\cal T}(\lambda) = {\cal L}(L|\lambda)\ {\cal L}(L-1|\lambda)\ \cdots\
	{\cal L}(1|\lambda)
   \label{monod}
\end{equation}
where $L$ is the number of sites in the lattice which we take to be even.
By construction this operator also satisfies a Yang Baxter equation
\begin{equation}
  R(\lambda,\mu) \left({\cal T}(\lambda)\otimes {\cal T(\mu)}\right)
  = \left({\cal T}(\mu)\otimes{\cal T}(\lambda)\right) R(\lambda,\mu) \ .
\label{intT}
\end{equation}

It might be interesting to point out that the entries fo the ${\cal
L}$-operator (\ref{lop}) form a representation of a quantum group: The
operators (we suppress the site index $n$)
\begin{eqnarray}
 &&  S^+ = {2\over i\sin\gamma\ m\Delta}\
      e^{ i\beta p/8}\ \rho\ e^{ i\beta p/8}\ , \nonumber \\[4pt]
 &&  S^- = {-2\over i\sin\gamma\ m\Delta}\
      e^{-i\beta p/8}\ \rho\ e^{-i\beta p/8}\ , \nonumber \\[4pt]
 &&   S^0 =  e^{-i\beta u/2}\ , \quad  S^1 =  e^{ i\beta u/2}\ \nonumber
\end{eqnarray}
satisfy the commutation relations of the quadratic (Sklyanin) algebra
\begin{eqnarray}
 &&	\left[ S^+, S^- \right] = {1\over{q-q^{-1}}} 
		\left((S^0)^2 - (S^1)^2\right)\ , \qquad
	\left[ S^0, S^1 \right] = 0\ , \nonumber \\[4pt]
 &&	S^\pm\ S^0 = q^{\mp1} S^0\ S^\pm\ , \qquad
	S^\pm\ S^1 = q^{\pm1} S^1\ S^\pm\ \nonumber
\end{eqnarray}
with $q=\exp(i\gamma)$.
For $q$ being a root of unity this algebra has finite dimensional cyclic
representations: for rational values of the parameter $\gamma/\pi=Q/P$ the
quantum operators entering the ${\cal L}$-operator can be written as
$2P\times2P$ matrices with elements
\[
   \chi = e^{i\beta u/2} \to \delta_{ab} e^{i\pi(a-1)/P}\ , \quad
   \pi = e^{i\beta p/4} \to \delta_{a+Q,b}\ , \qquad
   a,b=1,\ldots,2P\ , a+2P\equiv a\ .
\]

The definition of the ${\cal L}$ operator alone does not determine a
definite lattice model: In addition the Hamiltonian of the lattice \sG\
model needs to be specified.  For this choice there exist several different
possibilities (see \cite{izko:81,izko:82,ttf:84}). All of them are
completely integrable and can in fact be diagonalized simultaneously.
Furthermore all of them have the same continuum limit (\ref{sg:hamil}).
They differ from one another by higher orders in the lattice spacing
$\Delta$.
While all of them can be considered equivalently as a lattice
regularization of the continuum model we shall show below, how a
\emph{unique} lattice Hamiltonian can be chosen by requiring that it
has the ``same'' ground state wave function as the continuum model.  This
choice of the Hamiltonian will bring the dynamics of the lattice model as
close as possible to that of the continuum model.

\subsection{Algebraic \emph{Bethe Ansatz} for the lattice \sG\ model}
\label{sec:aba}
We shall consider the monodromy matrix (\ref{monod})
\begin{equation}
\label{monod2}
   {\cal T}(\lambda) = \left(\begin{array}{cc} A(\lambda) & B(\lambda) \\
      C(\lambda) & D(\lambda) \end{array} \right)\ .
\end{equation}
As a direct consequence of the Yang-Baxter equation (\ref{ybe}) for ${\cal
T}(\lambda)$ the trace of the monodromy matrix, the so-called transfer
matrix
\begin{equation}
   \tau(\lambda) = \hbox{\rm trace}\ {\cal T}(\lambda) 
                 = A(\lambda) + D(\lambda)
\label{transfm}
\end{equation}
commutes for different values of the spectral parameter $\lambda$, i.e.\
$\left[\tau(\lambda),\tau(\mu)\right] = 0$. Hence, it is the generator of
commuting integrals for the system which are diagonalized by the algebraic
\emph{Bethe Ansatz}. Starting point is the ``pseudo vacuum'' (or reference
state): To construct this simple eigenstate of $\tau(\lambda)$ we combine
the ${\cal L}$-operators in pairs:
\begin{equation}
   \widehat{\cal L}(n|\lambda) = {\cal L}(2n|\lambda)\ {\cal L}(2n-1|\lambda)
   \equiv \left( \begin{array}{cc} \alpha_n(\lambda) & \beta_n(\lambda) \\
      \gamma_n(\lambda) & \delta_n(\lambda) \end{array} \right)\ .
\label{pairlop}
\end{equation}
Choosing
\begin{equation}
   \langle u |0\rangle_n = \left\{ 1-
     2S\cos{\beta\over2}\left(u_{2n}+u_{2n-1}\right) \right\}^{-{1\over2}} 
   \delta\left( u_{2n} - u_{2n-1} -{\beta\over 4} + {2\pi\over\beta}\right)
\label{pseudov}
\end{equation}
(for rational $\gamma/\pi=Q/P$ the $\delta$-function can be replaced by a
Kronecker $\delta$-symbol and $|0\rangle_n$ will become normalizable) we
find from (\ref{pairlop})
\begin{eqnarray}
   \gamma_n(\lambda)\ |0\rangle_n &=& 0 \nonumber \\
   \alpha_n(\lambda)\ |0\rangle_n &=& 
       \left\{1+2S\cosh\left(2\lambda-i\gamma\right)\right\}|0\rangle_n
\label{id:loc}\\
   \delta_n(\lambda)\ |0\rangle_n &=& 
       \left\{1+2S\cosh\left(2\lambda+i\gamma\right)\right\}|0\rangle_n
   \nonumber
\end{eqnarray}
Now we can follow the standard steps of the algebraic \emph{Bethe Ansatz}.
As a consequence of (\ref{id:loc}) the ``global pseudo vacuum''
\begin{equation}
   |0\rangle = \prod_{n=1}^{L/2} |0\rangle_n
\label{gpseudov}
\end{equation}
is an eigenstate of the operators $A(\lambda)$ and $D(\lambda)$ (and hence
the transfer matrix (\ref{transfm})) with eigenvalues $a(\lambda)$ and
$d(\lambda)$, respectively:
\begin{equation}
   a(\lambda) 
     = \left\{1+2S\cosh\left(2\lambda-i\gamma\right)\right\}^{L\over2},
   \qquad
   d(\lambda)
     = \left\{1+2S\cosh\left(2\lambda+i\gamma\right)\right\}^{L\over2}\ .
\label{def:ad}
\end{equation}
More eigenfunctions of the transfer matrix are found by acting with the
operator $B(\lambda)$ on the pseudo vacuum
\begin{equation}
   \prod_{j=1}^N B(\lambda_j) |0\rangle
\label{bastate}
\end{equation}
provided that the $\{\lambda_j\}$ satisfy the \emph{Bethe Ansatz} equations
\begin{equation}
  \left( 
   {1+2S\cosh(2\lambda_j -i\gamma) \over 1+2S\cosh(2\lambda_j+i\gamma)}
  \right)^{L\over2} = - \prod_{k=1}^N
    {\sinh(\lambda_j-\lambda_k +i\gamma) \over 
     \sinh(\lambda_j-\lambda_k -i\gamma)}
\label{bae}
\end{equation}
The corresponding eigenvalue of the transfer matrix (\ref{transfm}) is
\begin{equation}
\label{tspec}
  \Lambda(\lambda|{\lambda_j}) =
   a(\lambda) \prod_{j=1}^N f(\lambda,\lambda_j)
   + d(\lambda) \prod_{j=1}^N f(\lambda_j,\lambda)
\end{equation}
where $f(\lambda,\mu)$ has been defined in (\ref{def:fg}).

The number $N$ of the Bethe Ansatz roots $\lambda_j$ can be identified with
the topological charge (\ref{defcon}). The correct lattice version in the
quantum case is
\begin{equation}
  \label{topchl}
  Q = {4\over\beta}\sum_{n=1}^{L/2} \left( u_{2n}-u_{2n-1} \right)
  + L {\pi-\gamma \over 2\gamma}\ .
\end{equation}
The difference in the coefficient compared to (\ref{defcon}) is related to
the fractional charge of the excitations. In \cite{kore:79} it was shown
that the fractional charge appears due to the repulsion beyond the cutoff
in the process of ultraviolet renormalization.  (\ref{topchl}) is the
number operator for particles
\[
   Q \prod_{j=1}^N B(\lambda_j) | 0\rangle =
       N \prod_{j=1}^N B(\lambda_j) | 0\rangle\ .
\]
One can prove that (here $\sigma^z$ is the Pauli matrix in the matrix space)
\[
   [Q,{\cal T}(\lambda)] = {1\over 2} [\sigma^z,{\cal T}(\lambda)]\ .
\]

Now we can discuss our choice of the lattice Hamiltonian for the lattice \sG\
model. As mentioned above we want to construct a lattice version resembling
the dynamics of the continuum model as closely as possible. According to
the standard quantization of the \sG\ model the ground state of
the continuum model contains no bound states (strings).

For possible lattice models we shall concentrate on the two integrable
models introduced in Refs.~\cite{izko:81,izko:82} and \cite{ttf:84}. The
latter has been constructed by Tarasov, Takhtajan and Faddeev (TTF) such
that it contains interactions of nearest neighbours on the lattice only. The
ground state for this Hamiltonian was found in Ref.~\cite{boiz:85}: in
addition to a Dirac sea of elementary particles it contains bound
states. In the continuum limit the density of the bound states vanishes,
thus reproducing the known results for the continuum model.  Apart from
the Hamiltonian the QISM yields higher integrals of motion. These
describe interactions over larger distances. Adding these interaction terms
to the TTF Hamiltonian with coefficients vanishing in the continuum limit
$\Delta\to0$ produces \emph{different} lattice Hamiltonians with the
\emph{same} continuum limit while preserving integrability. This is the
origin of the freedom in choice of the lattice hamiltonian.

Another Hamiltonian for the lattice \sG\ model has been introduced in
\cite{izko:81,izko:82}: The corresponding ground state for this Hamiltonian
has been constructed by Bogoliubov \cite{bogo:82}: He was able to prove
that in the interval $\pi/3 \le \gamma \le 2\pi/3$ the ground state is
built from elementary particles only---just as in the continuum model.
Furthermore, he found that the set of observable excitations coincides with
the continuum model. Hence, unlike the situation in the TTF model no phase
transition is met in performing the continuum limit. For the reasons
stated above we choose this Hamiltonian for our studies of correlation
functions.

It is given in terms of trace identities. Expressing the zeroes
$d(\kappa_\pm)=0$ and $a(\nu_\pm)=0$ of (\ref{def:ad}) as
\begin{equation}
  e^{2\kappa_\pm} = - b^{\pm1} e^{-i\gamma}\ , \quad
  e^{2\nu_\pm} =    - b^{\pm1} e^{ i\gamma}\ , \qquad
  \hbox{\rm where~}b = \frac{2S}{1+\sqrt{1-4S^2}} 
\label{def:kn}
\end{equation}
($\lambda_\pm={1\over2}\left(i\pi\pm\ln b\right)$ are the zeroes of the
quantum determinant (\ref{detqL}) of ${\cal L}$) the Hamiltonian of the
lattice \sG\ model considered here is given by
\begin{eqnarray}
\label{hamillsg}
   {\cal H}_{LSG}&=& -{m^2\Delta\over32b\sin\gamma} \left\{
	e^{i\gamma} \left( {\partial\over\partial\lambda} 
		\ln{\tau(\lambda)\over a(\lambda)}\right)_{\lambda=\kappa_+}
	-e^{-i\gamma} \left( {\partial\over\partial\lambda} 
		\ln{\tau(\lambda)\over a(\lambda)}\right)_{\lambda=\kappa_-}
	\right.	\nonumber \\
	&&\left.
	+e^{-i\gamma} \left( {\partial\over\partial\lambda} 
		\ln{\tau(\lambda)\over d(\lambda)}\right)_{\lambda=\nu_+}
	-e^{i\gamma} \left( {\partial\over\partial\lambda} 
		\ln{\tau(\lambda)\over d(\lambda)}\right)_{\lambda=\nu_-}
	\right\}
\end{eqnarray}
This is the model studied in \cite{izko:81,izko:82}. From (\ref{tspec}) one
finds that (\ref{bastate}) are eigenfunctions of this Hamiltonian with
energy eigenvalues given by
\begin{equation}
   {\cal H}_{LSG}\prod_{j=1}^N B(\lambda_j) | 0\rangle\ =
   \left( \sum_{k=1}^N h(\lambda_k) \right) 
         \prod_{j=1}^N B(\lambda_j) | 0\rangle\
\label{hspec}
\end{equation}
with the single particle energies
\begin{eqnarray}
  h(\lambda) = {m^2\Delta\over 32bi}\ \left\{
      {e^{i\gamma} \over\sinh(\kappa_+-\lambda)
			\sinh(\kappa_+-\lambda_j-i\gamma)}
     -{e^{-i\gamma}\over\sinh(\kappa_--\lambda)
			\sinh(\kappa_--\lambda-i\gamma)} \right.
\nonumber \\
   \left.
     -{e^{-i\gamma} \over\sinh(\nu_+-\lambda)
			\sinh(\nu_+-\lambda+i\gamma)}
     +{e^{i\gamma}\over\sinh(\nu_--\lambda_j)
			\sinh(\nu_--\lambda+i\gamma)}
  \right\}\ .
\label{dispsp}
\end{eqnarray}
In the continuum limit $\Delta\to0$ (which is reached by letting $b\to 0$
here) one immediately reproduces the result \cite{stf:79}
\[
   h(\lambda)\vert_{\Delta\to0} = {1\over 2} m^2\Delta \sin\gamma\ 
        \cosh 2\lambda
\]
for the single particle dispersion of the continuum model.

To find the solution of (\ref{bae}) corresponding to the ground state of
the model it is necessary to classify the possible configurations of
$\lambda_j$ in the complex plane according to the so called string
hypothesis \cite{tasu:72}. The details of this are not important in
the present context. It was found by Bogoliubov \cite{bogo:82} that
the ground state of (\ref{hamillsg}) is obtained by filling all
permitted states of pseudoparticles with rapidities $\lambda_j$ on the
line $\hbox{Im}\lambda=\pi/2$.  
Taking the logarithm of Eq.~(\ref{bae}) in such a state one obtains
\begin{equation}
  {L\over 2} p(\lambda_j+i{\pi\over2}) = 2\pi Q_j
     - i \sum_{k} \ln \left( {\sinh(\lambda_j-\lambda_k+i\gamma) \over
                              \sinh(\lambda_j-\lambda_k-i\gamma) } \right)\ .
\label{logbae}
\end{equation}
Here the $Q_j$ are distinct integers characterizing the state uniquely and
\[
   p(\lambda+i{\pi\over2}) = {-i} \ln \left(
        {1-2S\cosh(2\lambda-i\gamma) \over
                1-2S\cosh(2\lambda+i\gamma)} \right)\ .
\]
In the thermodynamic limit the density $\rho(\lambda_j)={1\over L}\partial
Q_j /\partial \lambda_j$ is then given in terms of the integral equation
\begin{equation}
   {1\over2}p'(\lambda+i{\pi\over2}) = 2\pi \rho(\lambda) 
     + \int_{-\infty}^{+\infty} d\mu K(\lambda-\mu) \rho(\mu)
\label{badens}
\end{equation}
where
\begin{equation}
\label{kkernel}
   K(\lambda) = {-\sin2\gamma
                 \over \sinh(\lambda+i\gamma)\sinh(\lambda-i\gamma)}
              = {-2\sin2\gamma\over{\cosh 2\lambda -\cos2\gamma}}\ .
\end{equation}
This integral equation can be solved by Fourier transform resulting in
($\Lambda = - \ln b$)
\begin{equation}
  \rho(\lambda) = {1\over4\pi} \int_{-\infty}^{\infty} dk\ e^{-ik\lambda}
    {\sinh{1\over2} k(\pi-\gamma)\over
	\sinh{1\over2}k\gamma\ \cosh{1\over2} k(\pi-\gamma)}
    \cos{1\over2} k\Lambda\ .
\end{equation}
Similarly one can compute the excitation energies. This is useful to find the
correct mass renormalization formula: To perform the continuum limit of the
\sG\ model one should let $\Delta\to0$ and simultaneously $m\to\infty$ as
\begin{equation}
	m = \hbox{const.}\ \Delta^{-\gamma/\pi}\ .
\label{sol_mass}
\end{equation}

\section{Algebraic formulation of correlation functions}
\label{sec:qcf}
For the evaluation of the correlation function (\ref{corrf}) we shall make
extensive use of the similarity (in the framework of the QISM) of the LSG
model with the spin-${1\over2}$ XXZ Heisenberg chain which is derived from a
monodromy matrix satisfying a Yang-Baxter equation with the same $R$-matrix
(\ref{rmat}) as the present model. The correlation functions corresponding to
(\ref{corrf}) in the XXZ model have have recently been studied in
\cite{efik:95,efik:95a}.

First we note, that the symmetry of the ${\cal L}$ operator (\ref{lsym})
implies for the ground state configuration consisting of rapidities
$\{\widetilde{\lambda_j}=\lambda_j+i\pi/2\}$ with real $\lambda_j$
\[
	\left( B(\widetilde{\lambda}_j)\right)^\dagger
      = C(\widetilde{\lambda}_j)\ .
\]
In order to express the correlation function (\ref{corrf}) in the
algebraic framework outlined above we first need to define the
lattice analog of the operator $Q(x)$ in (\ref{expaq}). The correct
expression is found to be
\begin{equation}
   Q_1(n) = {4\over\beta}\sum_{k=1}^{n/2}\left(u_{2k}-u_{2k-1}\right)
          + n \left({\pi-\gamma\over2\gamma}\right)
\end{equation}
which counts the number of particles in the interval $[1,n]$ ($n$ even). In
the continuum limit this expression becomes
\begin{equation}
   Q_1(n) \to {2\over\beta} \left( u(x)-u(0) \right) 
          + {x\over\Delta}\left({\pi-\gamma\over2\gamma}\right)\ ,
   \qquad x=n\Delta
\label{Qcont}
\end{equation}

Hence the lattice analog of the correlation function (\ref{corrf}) can
be written as 
\begin{equation}
\label{corrf2}
   \langle\Omega|\exp(\alpha Q_1(n))|\Omega\rangle \equiv
   \frac{\langle 0|\prod_{j=1}^{{N}}C(\widetilde{\lambda}_j)
         \exp(\alpha Q_1(n))
         \prod_{k=1}^{{N}} B(\widetilde{\lambda}_k)|0\rangle}
        {\langle 0|\prod_{j=1}^{{N}}C(\widetilde{\lambda}_j)
         \prod_{k=1}^{{N}} B(\widetilde{\lambda}_k)|0\rangle}
\end{equation}
where $\widetilde{\lambda}_j$ are solutions of the Bethe Ansatz equations
(\ref{bae}) for the ground state configuration.

Let us first study the norm appearing in the denominator of this
expression.  To evaluate this expression one should commute the
$C(\widetilde{\lambda_j})$ to the right of the product where they
annihilate the pseudovacuum $|0\rangle$.  Since the commutation
relations between the elements of the monodromy matrix (\ref{monod2})
are completely determined by the $R$-matrix we can use the result of
\cite{kore:82,efik:95} (see also \cite{gaudin2,gaudin3}) for the norm
of Bethe Ansatz states (after identifying $\gamma$ with $2(\pi-\eta)$
in paper \cite{efik:95}) 
\begin{equation}
   \langle 0 | \prod_{j=1}^N C(\widetilde{\lambda}_j)
	\prod_{j=1}^N B(\widetilde{\lambda}_k) | 0 \rangle
   = \left(-\sin\gamma\right)^N 
     \left\{\prod_{j\ne k} f(\lambda_j,\lambda_k)\right\}
     \left\{\prod_{j=1}^N a(\widetilde{\lambda}_j) 
		d(\widetilde{\lambda}_j)\right\}
     \det {\cal N}
\end{equation}
where the $N\times N$ matrix ${\cal N}$ is given by
\[
   {\cal N}_{jk} = \delta_{jk} 
     \left\{ i {\partial \over \partial\widetilde{\lambda}_j} 
                  \ln {a(\widetilde{\lambda}_j)\over d(\widetilde{\lambda}_j)}
	     + \sum_{n=1}^N K(\widetilde{\lambda}_j-\widetilde{\lambda}_n) 
     \right\}
     - K(\widetilde{\lambda}_j-\widetilde{\lambda}_k)\ .
\]
The functions $K(\lambda)$ and $a(\lambda)$, $d(\lambda)$ have been
introduced in the previous section. In the thermodynamic limit this
expression can be further simplified:
We rewrite ${\cal N} = {\cal I}\ \cdot {\cal J}$ where
\begin{eqnarray}
   && {\cal I}_{jk} = \delta_{jk} - \frac{K(\lambda_j-\lambda_k)}{\theta_k}\ ,
      \quad
      {\cal J}_{jk} = \delta_{jk} \theta_j \nonumber \\
   && \theta_j = i {\partial \over \partial\widetilde{\lambda}_j} 
                  \ln {a(\widetilde{\lambda}_j)\over d(\widetilde{\lambda}_j)}
	     + \sum_{n=1}^N K(\widetilde{\lambda}_j-\widetilde{\lambda}_n)\ .
  \nonumber
\end{eqnarray}
Comparing the last expression with Eqs.~(\ref{logbae}) and (\ref{badens})
for the ground state
density of particles one obtains $\theta_j = -2\pi L \rho(\lambda_j)$.
Performing the thermodynamic limit on the matrix ${\cal I}$ one finds that
it turns into a Fredholm integral operator $\widehat{\cal I}=1+
{1\over2\pi}\widehat{K}$ acting as
\begin{equation}
   \widehat{\cal I} * f \vert_\lambda = f(\lambda) + {1\over2\pi}
      \int_{-\infty}^{+\infty} d\mu K(\lambda-\mu) f(\mu)
\label{khat}
\end{equation}
Here $K(\lambda)$ is kernel given in (\ref{kkernel}).

Putting everything together we find
\begin{eqnarray}
   \langle 0 | \prod_{j=1}^N C(\widetilde{\lambda}_j)
	\prod_{j=1}^N B(\widetilde{\lambda}_k) | 0 \rangle
   &=& \left(2\pi L \sin\gamma \right)^N
     \left\{ \prod_{j\ne k} f(\lambda_j,\lambda_k) \right\} 
\nonumber \\
   && \times  \left\{ \prod_{j=1}^N a(\widetilde{\lambda}_j) 
                  d(\widetilde{\lambda}_j) \rho(\lambda_j) \right\}\ 
       \det \left( 1+ {1\over2\pi}\widehat{K} \right)\ .
\label{pro}
\end{eqnarray}

We now turn to the numerator of (\ref{corrf2}): to reduce the
evaluation of the expectation value of $\exp(\alpha Q_1(n))$ in a Bethe state
(\ref{bastate}) to the computation of scalar products we divide the
lattice of length $L$ into two sub-chains of length $n$ and $L-n$ and
associate a monodromy matrix with each of them, namely 
\begin{equation}
   {\cal T}(\lambda) = {\cal T}(2,\lambda){\cal T}(1,\lambda)\ ,
   \qquad
   {\cal T}(i,\lambda) = \left(\begin{array}{cc} 
		A_i(\lambda) & B(\lambda_i) \\
		C_i(\lambda) & D(\lambda_i) \end{array} \right)\ ,
   \qquad i=1,2\ .
\end{equation}
In terms of ${\cal L}$-operators they are given by
\begin{eqnarray}
   &&{\cal T}(2,\lambda) = {\cal L}(L,\lambda)\ {\cal L}(L-1,\lambda)\ \ldots\
			{\cal L}(n+1,\lambda) \nonumber \\
   &&{\cal T}(1,\lambda) = {\cal L}(n,\lambda)\ {\cal L}(n-1,\lambda)\ \ldots\
			{\cal L}(1,\lambda)\ . \nonumber
\end{eqnarray}
By construction these monodromy matrices satisfy the same Yang-Baxter
equation (\ref{intT}) as ${\cal T}(\lambda)$. Similarly, the global
reference state (\ref{gpseudov}) can be decomposed into a direct product of
pseudo vacua for the subchains $|0\rangle_2 \otimes |0\rangle_1$ (remember
that we have chosen $n$ to be even) which are eigenstates of $A_i(\lambda)$
and $D_i(\lambda)$
\begin{equation}
   A_i(\lambda)|0\rangle_i = a_i(\lambda)|0\rangle_i\ , \qquad
   D_i(\lambda)|0\rangle_i = d_i(\lambda)|0\rangle_i\\ ,
\label{adi}
\end{equation}
where $a_i(\lambda)$ and $d_i(\lambda)$ are given by (\ref{def:ad}) with
$L$ replaced by $n$ and $L-n$ for $i=1,2$, respectively. The creation and
annihilation operators $B_i(\lambda)$ and $C_i(\lambda)$ act
according to
\begin{equation}
   C_i(\lambda)|0\rangle_i = 0\ , \qquad
   \langle 0|B_i(\lambda)=0\\ .
\label{bci}
\end{equation}

In this decomposed quantum space the numerator of (\ref{corrf2}) can be
rewritten as (see e.g.\ \cite{vladb,efik:95})
\begin{eqnarray}
 &&\sum {_1}\langle0|\prod_{I_C} C_1(\widetilde{\lambda}^C_{I_C}) 
		     \prod_{{I_B}} B_1(\widetilde{\lambda}^B_{I_B})|0\rangle_1\
        {_2}\langle0|\prod_{II_C} C_2(\widetilde{\lambda}^C_{II_C}) 
		     \prod_{II_B} B_2(\widetilde{\lambda}^B_{II_B})|0\rangle_2\
 \nonumber \\
 && \quad  \times\ e^{\alpha n_1}
   \left\{\prod_{I_B,I_C} a_2(\widetilde{\lambda}^B_{I_B})
	d_2(\widetilde{\lambda}^C_{I_C})\right\}
   \left\{\prod_{II_B,II_C}a_1(\widetilde{\lambda}^C_{II_C})
	d_1(\widetilde{\lambda}^B_{II_B})\right\}
 \nonumber \\
 && \qquad \times\
   \left\{\prod_{I_B,II_B} f(\lambda^B_{I_B} ,\lambda^B_{II_B})\right\}
   \left\{\prod_{I_C,II_C} f(\lambda^C_{II_C} ,\lambda^C_{I_C})\right\}
\label{deco}
\end{eqnarray}
where the sum is over all partitions
\[
   \{\widetilde{\lambda}^B_{I_B}\}\cup \{\widetilde{\lambda}^B_{II_B}\} 
	= \{\widetilde{\lambda}\},\ 
   \{\widetilde{\lambda}^B_{I_B}\}\cap \{\widetilde{\lambda}^B_{II_B}\} 
	= \emptyset\ ,\
   \{\widetilde{\lambda}^C_{I_C}\}\cup \{\widetilde{\lambda}^C_{II_C}\} 
	= \{\widetilde{\lambda}\},\ 
   \{\widetilde{\lambda}^C_{I_C}\}\cap \{\widetilde{\lambda}^C_{II_C}\} 
	= \emptyset
\]
of the set $\{\widetilde{\lambda}\}$ with ${\rm card}\{
\widetilde{\lambda}_{I_B}\} = {\rm card}\{ \widetilde{\lambda}_{I_C}\}
= n_1$, ${\rm card}\{\widetilde{\lambda}_{II_C}\}= {\rm card}\{
\widetilde{\lambda}_{II_B}\}= N-n_1$. Due to \r{bci} we only
need to consider partitions such that the sizes of $I_B$ and
$I_C$ (and $II_B$ and $II_C$) are the same. We next turn to an
investigation of the scalar products occurring in \r{deco}. Owing to
\r{adi} and \r{bci} and the fact that the monodromy matrices ${\cal
T}(i,\lambda)$ fulfill the same Yang-Baxter equation (\ref{intT}) as
${\cal T}(\lambda)$ it is sufficient to consider scalar products on
the entire lattice
\[
{S}_N = {\langle 0| \prod_{j=1}^{N} C(\lambda_j^C)
\prod_{k=1}^{N}B(\lambda_k^B)|0\rangle}\ .
\]
Here we do not assume that the sets of spectral parameters
$\{\l^B\}$ and $\{\l^C\}$ are the same, and we also do not
impose the Bethe equations \r{bae}.
{}From \r{intT} and the action on the reference state $A(\lambda)\vac
=a(\lambda)\vac$, $D(\lambda)\vac =d(\lambda)\vac$ it follows that
scalar products can be represented as
\begin{equation}
{{S}_N = \sum_{A,D} \prod_{j=1}^N a(\l_j^A)\prod_{k=1}^N
d(\l_k^D) K_N\left(\matrix{
\{\l^C\}&\{\l^B\}\cr\{\l^A\}&\{\l^D\}\cr}\right)\ ,}
\label{sn}
\end{equation}
where the sum is over all partitions of $\{\l^C\}\cup \{\l^B\}$ into
two sets $\{\l^A\}$ and $\{\l^D\}$. The coefficients $K_N$ are
functions of the $\l_j$ and are {\sl completely determined by the
intertwining relation} \r{intT}. The $R$-matrix \r{rmat} is however
identical to the one for the spin$-\frac{1}{2}$ Heisenberg XXZ
model (after appropriate identifications of the coupling
constants). This implies that the coefficients $K_N$ for the
\sG\ model and the XXZ chain are identical, so that we can take
over the result for the XXZ case (see {\sl e.g.} \cite{efik:95}).
The main point is that the $K_N$'s can be represented as {\sl
determinants}. This is done in two steps: first the so-called {\sl
highest coefficients}, which are obtained for the partition
$\{\lambda^A\}=\{\lambda^C\}$, $\{\lambda^D\}=\{\lambda^B\}$, are
represented as determinants
\bea
K_N\left(\matrix{
\{\l^C\}&\{\l^B\}\cr\{\l^C\}&\{\l^B\}\cr}\right) &=& \left(\prod_{j>k}
g(\l_j^B, \l_k^B)g(\l_k^C, \l_j^C)\right) \prod_{j,k} h(\l_j^C,
\l_k^B) {\rm det}(M^{B}_{C})\ ,
   \nonumber \\
h(\mu,\nu) &=& {f(\mu,\nu)\over g(\mu,\nu)}\ ,\quad \left(M^B_C\right)_{jk}
= {g(\l_j^C,\l^B_k)\over h(\l_j^C,\l^B_k)} = t(\l_j^C,\l^B_k)\ .
\label{knh}
\eea
where from (\ref{def:fg})
\[
h(\l,\m)= {\sinh(\l-\m-i\gamma)\over -i\sin\gamma}\ ,\ 
t(\l,\m)= {-\sin^2\gamma \over \sinh(\l-\m-i\gamma)\sinh(\l-\m)} \ .
\]
In the second step arbitrary coefficients $K_N$ are then
expressed in terms of highest coefficients as follows
\bea
K_N\left(\matrix{
\{\l^C\}&\{\l^B\}\cr\{\l^A\}&\{\l^D\}\cr}\right) &=& \left(\prod_{j\in
AC}\prod_{k\in DC} f(\l_j^{AC}, \l_k^{DC})\right)\left(\prod_{l\in
AB}\prod_{m\in DB} f(\l_l^{AB}, \l_m^{DB})\right)\nn
&&\times\ K_n\left(\matrix{
\{\l^{AB}\}&\{\l^{DC}\}\cr\{\l^{AB}\}&\{\l^{DC}\}\cr}\right) 
K_{N-n}\left(\matrix{
\{\l^{AC}\}&\{\l^{DB}\}\cr\{\l^{AC}\}&\{\l^{DB}\}\cr}\right) \ .
\label{kna}
\eea
Using \r{knh} and \r{kna} in \r{sn} we obtain the following expression
for general scalar products in the lattice \sG\ model
\bea
S_N &=& \prod_{j>k}
g(\l_j^C,\l_k^C)g(\l_k^B,\l_j^B) \sum_{}{\rm sgn}(P_C){\rm
sgn}(P_B)\prod_{j,k} h(\l_j^{AB},\l_k^{DC})
\prod_{l,m} h(\l_l^{AC},\l_m^{DB})\nn
&&\times\ \prod_{l,k} h(\l_l^{AC},\l_k^{DC})\prod_{j,m}h(\l_j^{AB},\l_m^{DB})
{\rm det}(M^{AB}_{DC}){\rm det}(M^{AC}_{DB})\ ,
\label{Sn2}
\eea
where $P_C$ is the permutation $\{ \l_1^{AC},\ldots ,
\l_n^{AC},\l_1^{DC},\ldots ,\l^{DC}_{N-n}\}$ of $\{\l_1^C,\ldots
,\l_N^C\}$, $P_B$ is the permutation $\{ \l_1^{DB},\ldots ,
\l_n^{DB},\l_1^{AB},\ldots ,\l^{AB}_{N-n}\}$ of $\{\l_1^B,\ldots
,\l_N^B\}$, ${\rm sgn}(P)$ is the sign of the permutation $P$, and
\be
\left(M^{AB}_{DC}\right)_{jk} = t(\l_j^{AB},\l_k^{DC})
  d(\l_k^{DC})a(\l_j^{AB}) .
\ee
Following the steps first carried out in \cite{kore:87} it is now possible
to represent $S_N$ as a single determinant. The discussion for
\sG\ is identical to the only for the XXZ chain \cite{efik:95}
so that we only present a brief discussion of the necessary steps and
give the final result. We first note that the sum on the r.h.s. in
\r{Sn2} looks very similar to a Laplace decomposition of the
determinant of the {\sl sum} of two matrices
$(S_1)_{jk}=t(\l_j^C,\l_k^B) a(\l_j^C) d(\l_k^B)$ and
$(S_2)_{jk}=t(\l_k^B,\l_j^C) d(\l_j^C) a(\l_k^B)$ (see {\sl e.g.}
\cite{vladb} p. 221). However this does not reproduce the
$h(\l,\m)$-factors. This leads to the introduction of a {\sl dual quantum
field} $\varphi(\l)$ acting in a bosonic Fock space with vacua $\dv$
and $\nddv$\footnote{We use the same notation as in \cite{efik:95}.}
according to 
\bea
\varphi(\l) &=& p(\l) + q(\l),\ [\varphi(\l),\varphi(\m)]=0 \ ,\ \nddv
q(\l) = 0 = p(\l) \dv\ ,\nn 
{[ p(\l), q(\m)]} &=& -\ln(h(\l,\m)h(\m,\l))\ ,\ [p(\l),p(\m)]= 0
=[q(\l),q(\m)]\ .
\eea
We emphasize that the field $\varphi$ commutes for different
values of spectral parameters.
Using the dual field it is now possible to recast \r{Sn2} as a {\sl single}
determinant of the sum of two matrices
\bea
S_N &=& \prod_{j>k} g(\l_j^C,\l_k^C)
g(\l_k^B,\l_j^B) \prod_{j=1}^N
a(\l_j^C)d(\l_j^B)\prod_{j,k}h(\l_j^C,\l_k^B) \nddv \det S\dv\ ,\nn
S_{jk} &=& t(\l_j^C,\l_k^B) + t(\l_k^B,\l_j^C) {r(\l_k^B)\over r(\l_j^C)}
\exp\left(\varphi(\l_k^B)-\varphi(\l_j^C) \right)\nn
&&\qquad\qquad\times \prod_{m=1}^N {h(\l_k^B,\l_m^B)
h(\l_m^C,\l_j^C)\over h(\l_m^C,\l_k^B) h(\l_j^C,\l_m^B)} ,
\label{Sn3new}
\eea
where $r(\l) = {a(\l)\over d(\l)}$. The price we pay for representing
$S_N$ as a single determinant is the occurrence of the expectation
value in the dual space.

Using \r{Sn3new} in \r{deco} and then applying the dual field trick
several times it is possible to represent \r{deco} as a single
determinant of the sum of four matrices. This analysis is completely
analogous to the XXZ case treated in \cite{efik:95} so that we only
state the result:
\bea
&&\langle 0|\prod_{j=1}^{{N}}C(\widetilde{\lambda}_j)\exp(\alpha Q_1(n))
\prod_{k=1}^{{N}} B(\widetilde{\lambda}_k)|0\rangle =
     \left\{\prod_{j\ne k} f(\lambda_j,\lambda_k)\right\}
     \left\{\prod_{j=1}^N a(\widetilde{\lambda}_j) 
		d(\widetilde{\lambda}_j)\right\}
\nddv \det{{\cal G}}\dv\nn
{{\cal G}}_{jk}&=& t(\lt_j,\lt_k) + t(\lt_k,\lt_j) {r_1(\lt_j)\over
r_1(\lt_k)}\exp\left(\varphi_2(\lt_k) - \varphi_2(\lt_j)\right) \nn
&&+ \exp\left(\alpha +\varphi_4(\lt_k)-\varphi_3(\lt_j)\right)
\left[t(\lt_k,\lt_j) + t(\lt_j,\lt_k) {r_1(\lt_j)\over 
r_1(\lt_k)}\exp\left(\varphi_1(\lt_j) - \varphi_1(\lt_k)\right)\right]\nn
&&-i\ \delta_{jk}\sin\gamma\ \frac{\partial}{\partial\lt_j}\left(
\ln(r(\lt_j)) + \sum_{{n=1\atop n\neq j}}^N
\ln\left[\frac{h(\lt_j,\lt_n)}{h(\lt_n,\lt_j)}\right]\right),
\label{genfu6}
\eea
where $r_1(\l) = {a_1(\l)/ d_1(\l)}=\left(\frac{1+2S\cosh(2\l
-i\gamma)}{1+2S\cosh(2\l +i\gamma)}\right)^\frac{n}{2}$ and the
commuting dual fields $\varphi_a$ are defined according to 
\bea
   \varphi_a(\lambda) &=& p_a(\lambda) + q_a(\lambda)\ ,\ 
   \nddv q_a(\l) = 0 = p_a(\l) \dv \ , \nddv 0)=1\ ,\ a=1\ldots 4\ ,
   \nonumber \\[8pt]
   {[q_b(\mu), p_a(\lambda)]} &=&\!\!\!\!\left(
	\matrix{1&0&1&0\cr 0&1&0&1\cr 0&1&1&1\cr 1&0&1&1\cr}
   \right) \ln(h(\lambda ,\mu)) + \left(
	\matrix{1&0&0&1\cr 0&1&1&0\cr 1&0&1&1\cr 0&1&1&1\cr}
   \right) \ln(h(\mu, \lambda)), 
\label{dfcom}
\eea
where $a,b=1\ldots 4$. Here all terms not proportional to
$\delta_{jk}$ in ${\cal G}_{jk}$ 
are understood in the sense of l'Hospital for the diagonal elements.
In the thermodynamic limit further simplifications take
place. Following the analysis for the norms above we express ${\cal
G}$ as the product of two matrices ${\cal J}$ and ${\cal W}$ 
\be
   {\cal G}=-(\sin\gamma) {\cal W}{\cal J}\ , \quad 
   {\cal J}_{jk}=\delta_{jk}\theta_k\ ,\quad 
   {\cal W}_{jk} = \delta_{jk} -\frac{1}{\theta_k}{\cal V}(\lt_j,\lt_k)\ ,
\ee
where $\theta_j=-2\pi L \rho(\lambda_j)$ and
\bea
  (\sin\gamma) {\cal V}(\l ,\m)&=&t(\l,\m) + t(\m,\l) {r_1(\l)\over
  r_1(\m)}\exp\left(\varphi_2(\m) - \varphi_2(\l)\right) \nn
  +&&\hskip -1cm\exp\left(\alpha +\varphi_4(\m)-\varphi_3(\l)\right)
  \left[t(\m,\l) + t(\l,\m) {r_1(\l)\over 
  r_1(\m)}\exp\left(\varphi_1(\l) - \varphi_1(\m)\right)\right].
\label{v}
\eea
In the thermodynamic limit ${\cal W}$ turns into an integral operator
$\widehat{\cal W}=1+{1\over 2\pi}\widehat{V}$ acting as
\begin{equation}
   (1+\frac{1}{2\pi}\widehat{V}) * f \vert_\lambda = f(\lambda) + {1\over2\pi}
      \int_{-\infty}^{+\infty} d\mu V(\lambda,\mu) f(\mu)\ ,
\label{vhat}
\end{equation}
where the integral kernel is obtained from \r{v} as 
(the arguments of the dual fields are shifted by $i\pi/2$ which does
not alter the defining commutation relations \r{dfcom})
\bea
   &&V (\l, \m)  =  \frac{-\sin\gamma}{\sinh(\l-\m)} 
	\bigg\{ {1\over\sinh(\l-\m - i\gamma)} 
	 + {e_2^{-1} (\l) e_2(\m) \over\sinh(\l - \m + i\gamma)} \nn
	&&\qquad + \exp(\alpha +\varphi_4(\mu) - \varphi_3(\lambda))
	\bigg({1\over\sinh(\l-\m + i\gamma)}
	 + {e_1^{-1} (\mu) e_1(\lambda)\over\sinh(\lambda-\mu-i\gamma)}
	\bigg)
	\bigg\}\ ,  
\label{vcont}
\eea
with
\[
  e_2 (\l) = 
	\bigg( {1-2S\cosh(2\l + i\gamma) \over 1-2S\cosh(2\l - i\gamma)}
	\bigg)^{\frac{n}{2}} e^{\varphi_2 (\l )}\ ,
  \quad
  e_1 (\l) = 
	\bigg( {1-2S\cosh(2\l - i\gamma) \over 1-2S\cosh(2\l +i\gamma)}
	\bigg)^{\frac{n}{2}} e^{\varphi_1 (\l )} \ .
\]

Putting everything together we thus find
\be
\langle\Omega|\exp(\alpha Q_1(n))|\Omega\rangle \equiv
\frac{\nddv\det{\left(1+{1\over
2\pi}\widehat{V}\right)}\dv}{\det{\left(1+\frac{1}{2\pi}\widehat{K}\right)}}\
, 
\label{final}
\ee
where $1+{1\over 2\pi}\widehat{V}$ and $1+\frac{1}{2\pi}\widehat{K}$
are integral operators acting according to \r{vhat} and \r{khat} with
kernels defined in \r{kkernel} and \r{vcont}.

\section{Continuum linit}
\label{sec:cont}
As mentioned in the introduction the purpose of the present work is to
determine correlators for the SG Quantum Field Theory, and the lattice
model studied above is used merely as a regulator for the UV
divergences. We are therefore interested in the {\sl continuum limit}
of the determinant representation \r{final}. As mentioned above the SG
Quantum Field Theory is recovered from the lattice regularization by
taking the lattice spacing to zero $\Delta\to 0$ and simultaneously
the bare mass $m$ to infinity keeping $m \Delta^{\frac{\g}{\pi}}$
fixed \cite{bogo:82}.
In order to take the continuum limit we now employ the following
regularisation for the integral operators in (\ref{final}): we
restrict the integration for the integral operator
$1+{1\over{2\pi}}\widehat{V}$ to the interval $[-\La,\La]$, and then
take $\Delta\rightarrow 0$ in such a way that $S\cosh(2\l)\ll 1\
\forall\l\in[-\La,\La]$ (recall \r{defrho} for the relation of $S$ and
$\Delta$). Using this regularisation the $e_j(\l)$'s simplify to 
\begin{equation}
   e_2(\l)
	= \exp(-ip\sinh(2\l)+\varphi_2(\l))\ ,\quad 
   e_1(\l) = \exp(ip\sinh(2\l)+\varphi_1(\l))\ , 
\label{e_cont}
\end{equation}
where 
\be
p=\frac{c^2}{8}\Delta^{\frac{\pi-2\g}{\pi}}\sin(\g) n\Delta\ .
\label{p}
\ee
Here we have used (\ref{sol_mass}) and $n\Delta=x$ should be
identified with the continuum distance. The constant $c$ is given in
terms of the physical soliton mass.

This regularisation allows to embed the determinant (\ref{final}) into a
system of integrable integro-differential equations which we shall need
later to determine the subleading terms in the asymptotic expansion of the
correlation functions: with (\ref{e_cont}) the kernel (\ref{vcont}) can be
brought into standard form \cite{vladb}:
We perform a change of variables $z=\exp(2\l)$, and replace the factors
$(\sinh(\l-\m\pm i\g))^{-1}$ in \r{vcont} by an integration over an
exponential. Then the transpose of the kernel \r{vcont} reads (up to a
similarity transform which leaves the determinant unchanged)
\begin{equation}
	\frac{1}{2\pi}V^T(z_1,z_2)=
		\frac{i}{z_1-z_2}\int_0^\infty ds\sum_{j=1}^4
			E_j(z_2|s)e_j(z_1|s) \ ,
\label{stand}
\end{equation}
where
\bea
   e_1(z|s)&=& \frac{\kappa}{\sqrt{2\pi}}
	\exp(\varphi_4(z))|2,z,s\rangle\ ,\quad
   E_1(z|s)= -\frac{\kappa}{\sqrt{2\pi}}
	\exp(-\varphi_3(z))\langle 2,z,s|\nn
   e_2(z|s)&=& \frac{1}{\sqrt{2\pi}}| 1,z,s\rangle\ ,\quad
   E_2(z|s)= \frac{1}{\sqrt{2\pi}} \langle1,z,s|\nn
   e_3(z|s)&=& \frac{1}{\sqrt{2\pi}}
	\exp(-ipk(z)+\varphi_2(z))|2,z,s\rangle\ ,\quad
   E_3(z|s)= -\frac{1}{\sqrt{2\pi}}
	\exp(ipk(z)-\varphi_2(z))\langle 2,z,s|\nn
   e_4(z|s)&=&\frac{\kappa}{\sqrt{2\pi}}
	\exp(-ipk(z)-\varphi_1(z)+\varphi_4(z))|1,z,s\rangle\,\nn
   E_4(z|s)&=&\frac{\kappa}{\sqrt{2\pi}}
	\exp(ipk(z)+\varphi_1(z)-\varphi_3(z))\langle 1,z,s| .
\label{def:eE}
\eea
Here we use the notation $k(z)=\frac{1}{2}(z-z^{-1})$, $w=\exp(i\g)$,
$\kappa=\exp(\frac{\alpha}{2})$, and
\begin{equation}
   |1,z,s\rangle = \sqrt{2z\sin(\g)} \exp(izws) = \langle 2,z,s|\ ,\qquad
   |2,z,s\rangle = \sqrt{2z\sin(\g)} \exp(-i\frac{z}{w}s) = \langle 1,z,s|
\label{s-proj}
\end{equation}
are normalized in such a way that
$\langle 1|1\rangle = \int_0^\infty ds\ \langle 1,z,s|1,z,s\rangle =1$,
and similarly $\langle 2|2\rangle =1$.


The inverse of the integral operator $1+\frac{1}{2\pi}\widehat{V}^T$ is
defined by
\begin{eqnarray}
	(1-\widehat{R})*(1+\frac{1}{2\pi}\widehat{V}^T)=1
	=(1+\frac{1}{2\pi}\widehat{V}^T)*(1-\widehat{R})\ ,\nn
	\widehat{R}=(1+\frac{1}{2\pi}\widehat{V}^T)^{-1}
	*\frac{1}{2\pi}\widehat{V}^T\ .
\label{res}
\end{eqnarray}
In terms of the functions $f_j(z|s), F_j(z|s)$
\be
(1-\widehat{R})*e_j\bigg|_{z,s} = f_j(z|s)\ ,\qquad
E_j*(1-\widehat{R})\bigg|_{z,s} = F_j(z|s)\ .
\label{deff}
\ee
the kernel of $\widehat{R}$ can be written in a form similar to \r{stand}
\begin{equation}
	R(z_1,z_2) = \frac{i}{z_1-z_2}\sum_{j=1}^4
		\int_0^\infty ds\ f_j(z_1|s)F_j(z_2|s) \ .
\label{resker}
\end{equation}
as can be seen by acting with $(1+\frac{1}{2\pi}\widehat{V}^T)$ on \r{resker}.
\section{Integro-Differential equations}
\label{sec:ide}
Let us now derive integro-differential equations determining the functions
$f_j(z|s)$ and $F_j(z|s)$. The analog of these equations in the case
of impenetrable bosons proved very useful for the anlysis of the
corresponding Riemann-Hilbert problem and we expect the equations
below to play a similar role for the problem at hand. To this end we
consider derivatives with respect to $p$ and the integration boundary
$\La$.  For the $\La$-derivatives we 
find
\begin{eqnarray}
   \partial_\La f_j(z|s)
	+\sum_{l=1}^4\int_0^\infty dt U_{jl}(z|s,t)f_l(z|t) &=& 0\ ,\nn 
   \partial_\La F_j(z|s)
	-\sum_{l=1}^4\int_0^\infty dt F_l(z|t)U_{lj}(z|t,s) &=& 0\ ,
\label{dLf}
\end{eqnarray}
where
\be
U_{jk}(z|s,t)=\frac{2ie^{2\La}}{z-e^{2\La}}\ f_j(e^{2\La}|s)\
F_k(e^{2\La}|t)+\frac{2ie^{-2\La}}{z-e^{-2\La}}\ f_j(e^{-2\La}|s) \
F_k(e^{-2\La}|t)\ .
\ee
The $p$-derivatives of the functions $f_j(z|s)$ obey the
integro-differential equations
\bea
\partial_p f_j(z|s)&=&\left(-ik(z)f_j(z|s)+\frac{1}{2}\sum_{l=1}^4
\left[B^{(0)}_{jl}+\frac{1}{z}B^{(1)}_{jl}\right]*f_l\bigg|_{z,s}\right)
(\delta_{j,3}+\delta_{j,4})\nn
&&-\frac{1}{2z}\sum_{k=3}^4C^{(1)}_{jk}*\sum_{l=1}^4\left[I-i
B^{(1)}\right]_{kl}*f_l\bigg|_{z,s}-\frac{1}{2}\sum_{k=3}^4
C^{(0)}_{jk}*f_k\bigg|_{z,s}\ ,
\label{dpf}
\eea
where $I_{jk}(s,t)=\delta_{jk}\delta(s-t)$ and where the integral
operators $B^{(n)}$ and $C^{(n)}$ are defined as 
\bea
B^{(n)}_{jk}(s,t)&=&\int_\A^\B \frac{dz}{z^n}e_j(z|s)F_k(z|t)\ ,\nn
C^{(n)}_{jk}(s,t)&=&\int_\A^\B \frac{dz}{z^n}f_j(z|s)E_k(z|t)\ .
\label{pot}
\eea
We note the following relations between the integral operators
$B^{(n)}_{jk}$ and $C^{(n)}_{jk}$
\be
B^{(0)}_{jk}(s,t) = C^{(0)}_{jk}(s,t)\ ,\qquad
[I-iB^{(1)}]_{jk}*[I+iC^{(1)}]_{kl}\bigg|_{s,t}= \delta_{jl}\delta(s-t) .
\label{inv}
\ee
These identities can be easily proved by using \r{deff}. From now on
we will replace $B^{(0)}_{jk}$ in all expressions by $C^{(0)}_{jk}$.
The IDE for $F_j(z|s)$ are found to be
\bea
\partial_p F_j(z|s)&=&\left(ik(z)F_j(z|s)+\frac{1}{2}\sum_{l=1}^4
F_l*\left[C^{(0)}_{lj}+\frac{1}{z}C^{(1)}_{lj}\right]\bigg|_{z,s}\right)
(\delta_{j,3}+\delta_{j,4})\nn
&&-\frac{1}{2z}\sum_{k=3}^4\sum_{l=1}^4F_l*[I+iC^{(1)}]_{lk}*
B^{(1)}_{kj}\bigg|_{z,s}-\frac{1}{2}\sum_{k=3}^4
F_k*C^{(0)}_{kj}\bigg|_{z,s}\ ,
\label{dpF}
\eea

The ``potentials'' $B^{(n)}$ and $C^{(n)}$ obey the equations
\bea
\partial_p C^{(n)}_{jk}(s,t)&=&-\frac{1}{2}\sum_{m=3}^4 
C^{(0)}_{jm}*C^{(n)}_{mk}\bigg|_{s,t}
-\frac{1}{2}\sum_{m=3}^4
C^{(1)}_{jm}*\sum_{l=1}^4\left[I-iB^{(1)}\right]_{ml}*
C^{(n+1)}_{lk}\bigg|_{s,t}\nn
&-&\frac{i}{2}(\delta_{j,3}+\delta_{j,4})\left[C_{jk}^{(n-1)}(s,t)-
C_{jk}^{(n+1)}(s,t)+i\sum_{l=1}^4 C^{(0)}_{jl}*C^{(n)}_{lk}\bigg|_{s,t}
+B^{(1)}_{jl}*C^{(n+1)}_{lk}\bigg|_{s,t}\right]\nn
&+&\frac{i}{2}(\delta_{k,3}+\delta_{k,4})\left[C_{jk}^{(n-1)}(s,t)-
C_{jk}^{(n+1)}(s,t)\right],
\eea

\bea
\partial_p B^{(n)}_{jk}(s,t)&=&-\frac{1}{2}\sum_{m=3}^4 
B^{(n)}_{jm}*C^{(0)}_{mk}\bigg|_{s,t}
-\frac{1}{2}\sum_{m=3}^4\sum_{l=1}^4
B^{(n+1)}_{jl}*\left[I+iC^{(1)}\right]_{lm}*
B^{(1)}_{mk}\bigg|_{s,t}\nn
&+&\frac{i}{2}(\delta_{k,3}+\delta_{k,4})\left[B_{jk}^{(n-1)}(s,t)-
B_{jk}^{(n+1)}(s,t)-i\sum_{l=1}^4 B^{(n)}_{jl}*C^{(0)}_{lk}\bigg|_{s,t}
+B^{(n+1)}_{jl}*C^{(1)}_{lk}\bigg|_{s,t}\right]\nn
&-&\frac{i}{2}(\delta_{j,3}+\delta_{j,4})\left[B_{jk}^{(n-1)}(s,t)-
B_{jk}^{(n+1)}(s,t)\right].
\eea

The derivatives with respect to $\La$ are given by
\bea
\partial_\La C^{(0)}_{jk}(s,t)&=&
2e^{2\La}f_j(e^{2\La}|s)F_k(e^{2\La}|t)+
2e^{-2\La}f_j(e^{-2\La}|s)F_k(e^{-2\La}|t)\ ,\nn
\partial_\La C^{(1)}_{jk}(s,t)&=&
2f_j(e^{2\La}|s)\left(F_k(e^{2\La}|t)
+iF_l*C^{(1)}_{lk}\bigg|_{e^{2\La},t}\right)+\La\rightarrow -\La\nn
\partial_\La C^{(2)}_{jk}(s,t)&=&
2e^{-2\La}f_j(e^{2\La}|s)\left(F_k(e^{2\La}|t)
+iF_l*C^{(1)}_{lk}\bigg|_{e^{2\La},t}+ie^{2\La}F_l*C^{(2)}_{lk}
\bigg|_{e^{2\La},t}\right)+\La\rightarrow -\La\ , \nn
\partial_\La B^{(1)}_{jk}(s,t)&=&
2F_k(e^{2\La}|t)\left(f_j(e^{2\La}|s)
-iB^{(1)}_{jl}*f_l\bigg|_{e^{2\La},s}\right)+\La\rightarrow -\La\ .
\eea

Eqs.~(\ref{dLf}), (\ref{dpf}) and (\ref{dpF}) form a Lax pair. Their
consistency is implied by the following relation for the cross-derivatives
\begin{equation}
   \partial_p\partial_\La f_j(z|s)=\partial_\La\partial_p f_j(z|s)\ .
\end{equation}
In order to simplify the computations we first introduce some
notation. We rewrite \r{dpf} as
\be
\partial_p f_j(z|s)=
-\frac{i}{2}zf_j(z|s)(\delta_{j,3}+\delta_{j,4})
+\sum_{l=1}^4a_{jl}*f_l\bigg|_{z,s}
+\frac{1}{z}\sum_{l=1}^4b_{jl}*f_l\bigg|_{z,s}\ ,
\ee
where 
\bea
a_{jl}(s,t)&=& \frac{1}{2}C^{(0)}_{jl}\left(\delta_{j,3}+\delta_{j,4}- 
\delta_{l,3}-\delta_{l,4}\right)\ ,\nn
b_{jl}(s,t)&=& \frac{i}{2}\delta_{jl}\delta(s-t)
+\frac{1}{2}B^{(1)}_{jl}(\delta_{j,3}+\delta_{j,4})-
\frac{1}{2}\sum_{k=3}^4 C^{(1)}_{jk}[I-iB^{(1)}]_{kl}\ .
\eea
In the same notation (\ref{dpF}) can be written as
\begin{equation}
  \partial_p F_j(z|s) =
  {i\over2}z F_j(z|s) (\delta_{j,3}+\delta_{j,4})
  - \sum_{l=1}^4 F_l*a_{lj}\bigg|_{z,s}
  - {1\over z} \sum_{l=1}^4 F_l*b_{lj}\bigg|_{z,s}\ .
\end{equation}

Similarly we introduce the notation
\be
U_{jl}(z|s,t) = \frac{A_{jl}(\La|s,t)}{z-e^{2\La}} +
\frac{A_{jl}(-\La|s,t)}{z-e^{-2\La}}\ ,
\ee
where $A_{jk}(\La|s,t)=2ie^{2\La}\ f_j(e^{2\La}|s)\ F_k(e^{2\La}|t)$. In
what follows we will denote by $\widehat{A}_{jk}(\La)$ the integral
operator in the $s$-variable with kernel $A(\La|s,t)$.
After some calculations we arrive at the following equations
\bea
\partial_\La\partial_p f_j(z|s) &=&
\frac{iz}{2}(\delta_{j,3}+\delta_{j,4})\left(
\sum_{m=1}^4\frac{\wh{A}(\La)_{jm}*f_m\bigg|_{z,s}}{z-e^{2\La}}+\La\rightarrow
-\La\right)\nn
&+&\sum_{m=1}^4 \partial_\La a_{jm}*f_m\bigg|_{z,s}
-\sum_{l=1}^4 a_{jl}*\left(\sum_{m=1}^4\frac{\wh{A}_{lm}(\La)*f_m}{z-e^{2\La}}
+\La\rightarrow -\La\right)\bigg|_{z,s}\nn
&+&\frac{1}{z}\sum_{m=1}^4 \partial_\La b_{jm}*f_m\bigg|_{z,s}
-\frac{1}{z}\sum_{l=1}^4
b_{jl}*\left(\sum_{m=1}^4\frac{\wh{A}_{lm}(\La)*f_m}{z-e^{2\La}}
+\La\rightarrow -\La\right)\bigg|_{z,s}\ ,
\label{crossl}
\eea

\bea
\partial_p\partial_\La f_j(z|s) &=&
-\sum_{m=1}^4\left(\frac{\partial_p\wh{A}(\La)_{jm}*f_m\bigg|_{z,s}}
{z-e^{2\La}}+\La\rightarrow -\La\right)\nn
&+&\frac{iz}{2}\sum_{m=1}^4\left(
\frac{\wh{A}(\La)_{jm}*f_m\bigg|_{z,s}}{z-e^{2\La}} +\La\rightarrow
-\La\right)(\delta_{m,3}+\delta_{m,4})\nn 
&-&\sum_{l=1}^4\sum_{m=1}^4\left( \frac{\wh{A}(\La)_{jl}}{z-e^{2\La}}
+\La\rightarrow
-\La\right)*\left(a_{lm}*f_m+\frac{1}{z}b_{lm}*f_m\right)\bigg|_{z,s}\
.
\label{crossr}
\eea
In order to equate \r{crossl} and \r{crossr} we first rewrite both
equations in the form $O_{jm}*f_m$, where $O$ are complicated
integral operators, and then ``truncate'' the $f_m$'s from the
resulting expressions, which amounts to supposing that they form an
independent set of functions in the space the integral operators act
in. In the next step we then compare the resulting expressions (which
are both meromorphic functions of $z$) at the singular points
$z=\infty,0,e^{\pm 2\La}$. If they (their residues) are equal at these
points the expressions coincide for all values of $z$.
For $z\rightarrow\infty$ we get the condition
\be
\partial_\La a_{jm}(s,t) = \frac{i}{2}(\delta_{m,3}+\delta_{m,4}-
\delta_{j,3}-\delta_{j,4})(A_{jm}(\La|s,t)+A_{jm}(-\La|s,t))\ ,
\ee
At $z=0$ we obtain
\be
\partial_\La b_{jm}(s,t)=
\sum_{l=1}^4
\left(e^{-2\La}\left[\wh{A}_{jl}(\La)*b_{lm}-b_{jl}*\wh{A}_{lm}(\La)\right]
+\La\rightarrow -\La\right)\bigg|_{s,t}\ .
\ee
At $z=e^{2\La}$ we get
\bea
\partial_p A_{jm}(\La|s,t)&=&
\frac{i}{2}e^{2\La}(\delta_{m,3}+\delta_{m,4}-\delta_{j,3}-\delta_{j,4})
A_{jm}(\La|s,t)\nn
&+&\sum_{l=1}^4\left( [a_{jl}+e^{-2\La}b_{jl}]*\wh{A}_{lm}(\La)
-\wh{A}_{jl}(\La)*[a_{lm}+e^{-2\La}b_{lm}]\right)\bigg|_{s,t}\ ,
\label{exp2l}
\eea
whereas the condition from $z=e^{-2\La}$ is obtained by taking
$\La\rightarrow -\La$ in \r{exp2l}. 
It is straightforward to show that these equations hold by inserting the
expressions for $a$, $b$ and $A$ and using the identities for the $p$- and
$\La$-derivatives of $C^{(n)}_{jk}$ and $B^{(n)}_{jk}$ written above.

Finally, to relate the functional determinant in \r{final} to the
quantities introduced above we turn to the logarithmic derivatives of
$\det(1+\frac{1}{2\pi}\widehat{V}^T)$.

The derivative with respect to $p$ is given by
\begin{equation}
  \partial_p\ln\left(\det(1+\frac{1}{2\pi}\widehat{V}^T)\right)
	={\rm tr}\left((1-\widehat{R})
		*\frac{1}{2\pi}\partial_p\widehat{V}^T\right).
\label{logder}
\end{equation}
Using \r{def:eE} we find that
\bea
\frac{1}{2\pi}\partial_p{V}^T(z_1,z_2)&=&
\frac{k(z_1)-k(z_2)}{z_1-z_2}\int_0^\infty ds\ \sum_{j=3}^4
e_j(z_1|s)E_j(z_2|s)\nn
&=&\frac{1}{2}(1+\frac{1}{z_1z_2})\int_0^\infty ds\ \sum_{j=3}^4
e_j(z_1|s)E_j(z_2|s).
\eea
This implies that
\bea
(1-\widehat{R})*\frac{1}{2\pi}\partial_p\widehat{V}^T\bigg|_{z_1,z_2}
&=&\frac{1}{2}\int_0^\infty ds\ \sum_{j=3}^4\int_\A^\B dz\left[
\delta(z_1-z)-R(z_1,z)\right]\frac{e_j(z|s)E_j(z_2|s)}{zz_2} \nn
&&+\frac{1}{2}\int_0^\infty ds\ \sum_{j=3}^4 f_j(z_1|s)E_j(z_2|s).
\label{one}
\eea
Using the representation \r{resker} of $R(z_1,z_2)$ we rewrite the r.h.s. as
\begin{eqnarray}
{\rm r.h.s}&=&\frac{1}{2}\int_0^\infty ds\sum_{k=3}^4 \int_\A^\B dz
\left[\delta(z_1-z)-R(z_1,z)\right]\frac{e_k(z|s)E_k(z_2|s)}{z_1z_2}\nn
&&-\frac{i}{2}\int_0^\infty ds\int_0^\infty dt\sum_{k=3}^4\sum_{l=1}^4
\int_\A^\B dz\
\frac{f_l(z_1|t)F_l(z|t)}{z_1z}\frac{e_k(z|s)E_k(z_2|s)}{z_2} \nn
&&+\frac{1}{2}\int_0^\infty ds\ \sum_{j=3}^4 f_j(z_1|s)E_j(z_2|s)\ .
\label{rhs}
\end{eqnarray}
Using this with \r{pot} in \r{logder} we finally arrive at
\begin{equation}
   \partial_p\ln\left(\det(1+\frac{1}{2\pi}\widehat{V}^T)\right)=
	\frac{1}{2}\sum_{k=3}^4\int_0^\infty ds 
	\left[C^{(0)}_{kk}(s,s)+C^{(2)}_{kk}(s,s)
		-i\sum_{l=1}^4B^{(1)}_{kl}*C^{(2)}_{lk}\bigg|_{s,s}
	\right].
\label{logder2}
\end{equation}

The logarithmic derivative of the determinant with respect to $\La$ is 
\begin{equation}
	\partial_\La \ln\left(\det(1+\frac{1}{2\pi}\widehat{V}^T)\right) =
	2e^{2\La} R(e^{2\La},e^{2\La})+2e^{-2\La}R(e^{-2\La},e^{-2\La})\ .
\end{equation}
After some manipulations similar to the case of the Bose gas (see
\cite{vladb}) this can be rewritten as
\begin{eqnarray}
&&R(e^{2\La},e^{2\La})=\frac{ie^{-2\La}}{2}\sum_{j=1}^4\int_0^\infty ds\
	F_j(e^{2\La}|s)\frac{d}{d\La}f_j(e^{2\La}|s)\nn
&&\qquad-\frac{e^{-4\La}}{2\sinh(2\La)}\left[\sum_{j=1}^4\int_0^\infty ds\
	f_j(e^{-2\La}|s) F_j(e^{2\La}|s)\right]\left[\sum_{l=1}^4 \int_0^\infty dt\
	f_l(e^{2\La}|t) F_l(e^{-2\La}|t)\right]\nn
&&R(e^{-2\La},e^{-2\La})=-\frac{ie^{2\La}}{2}\sum_{j=1}^4 \int_0^\infty ds\
	F_j(e^{-2\La}|s)\frac{d}{d\La}f_j(e^{-2\La}|s)\nn
&&\qquad-\frac{e^{4\La}}{2\sinh(2\La)}\left[\sum_{j=1}^4\int_0^\infty ds\
	f_j(e^{2\La}|s) F_j(e^{-2\La}|s)\right]\left[\sum_{l=1}^4 \int_0^\infty dt\
	f_l(e^{-2\La}|t) F_l(e^{2\La}|t)\right]. 
\end{eqnarray}
This embeds the determinant into the system of integrable integro-differential
equations derived above.

\section{The Riemann-Hilbert Problem}
\label{sec:rhp}
\begin{figure}[h]
	\begin{center}
	\leavevmode
	\epsfxsize=0.5\textwidth
	\epsfbox{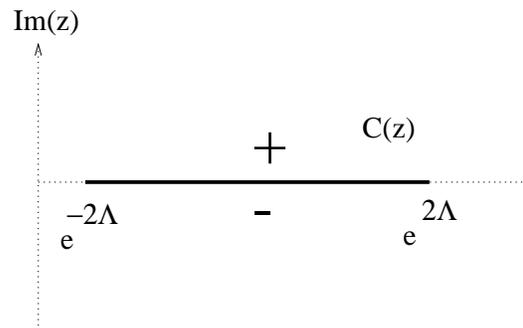}
	\end{center}
	\caption{Conjugation Contour for the RHP}
	\label{fig:RHP1}
\end{figure}

In this section we show that the results of the previous section can be
reformulated in terms of an infinite-dimensional Riemann-Hilbert problem
(RHP) for an integral operator valued function $Y(z)$. This connection
will enable us to determine the asymptotic behaviour of the
correlation function. We introduce the conjugation matrix $G(z)$
of this Riemann-Hilbert problem as
\be
\left[G(z|s,t)\right]_{ij}=\delta_{ij}\delta(s-t)+2\pi e_i(z|s)E_j(z|t)\ .
\label{cm}
\ee
It's elements can be expressed in terms of the projectors \r{s-proj}, e.g.
\bea
\left[G(z|s,t)\right]_{11}&=&
\delta(s-t)-\k^2 \exp(\varphi_4(z)-\varphi_3(z))|2,z,s\rangle\langle 2,z,t| \nn
\left[G(z|s,t)\right]_{12}&=&
\k \exp(\varphi_4(z))|2,z,s\rangle\langle 1,z,t| \nn
\ldots&&\nonumber
\eea
Consider now an integral-operator valued function $Y(z)$ with kernel
$Y_{jk}(z|s,t)$, $j,k=1,\ldots,4$, $s,t\in [0,\infty)$ acting on a vector
$f$ of functions of $z$ and $s$ according to
\begin{equation}
	[Y(z)*f(z)]_j=\int_0^\infty dt \sum_{k=1}^4
	Y_{jk}(z|s,t)f_k(z|t)\ .
\end{equation}
$Y(z)$ is solution to the following RHP
\begin{itemize}
\item $Y(z)= I +\sum_{k=1}^\infty \frac{M_k}{z^k}\ {\rm for}\
  z\rightarrow\infty$.
\item $Y(z)$ is analytic throughout the complex plane with the exception of
  the contour $C$, which is the interval $[\exp(-2\La),\exp(2\La)]$ on
  the real axis (see Fig.~\ref{fig:RHP1}).
\item $Y^-(z)= Y^+(z)G(z)$ on $C$ where $Y^\pm(z)$ are the boundary values
  as indicated in Fig.~\ref{fig:RHP1} and $G(z)$ is the conjugation matrix
  \r{cm}.
\end{itemize}
This RHP can be rewitten as the system of singular integral equations
\be
Y^+(z)= I+\frac{1}{2\pi i}\int_{-\infty}^\infty dz^\prime\
\frac{Y^+(z^\prime)[I-G(z^\prime)]}{z^\prime-z-i0}\ .
\label{inteq}
\ee
The solution of \r{inteq} can be expressed in terms of the functions
$E$ and $f$ defined in Section \ref{sec:cont} as
\be
  Y_{ij}^+(z|s,t)= \delta_{ij}\delta(s-t)+i\int_\A^\B dz^\prime
  \frac{f_i(z^\prime|s)E_j(z^\prime|t)}{z^\prime-z-i0}
\ee
which follows from the identity
\be
f_j(z|s)=\int_0^\infty dt\ Y_{jk}(z|s,t)\ e_j(z|t)\ .
\ee


The potentials $B^{(1)}$ and $C^{(n)}$ \r{pot} can be related to the
solution $Y(z)$ of the RHP through asymptotic expansions around $0$
and $\infty$. We find 
\begin{eqnarray}
	&&Y_{jk}(z)\longrightarrow I+iC^{(1)}+izC^{(2)}+iz^2C^{(3)}
		+{\cal O}(z^3) \qquad {\rm for}\ z\rightarrow 0\ ,
\label{contr1}\\
	&&Y_{jk}(z)\longrightarrow I-\frac{i}{z}C^{(0)}
		+{\cal O}(z^{-2})\qquad {\rm for}\ z\rightarrow\infty .
\label{contr2}
\end{eqnarray}
{}From \r{contr1} and \r{inv} we find
\be
[I-iB^{(1)}]*C^{(2)}= -i(Y^{-1}(0)\frac{d}{dz}\bigg|_{z=0}Y(z))\ .
\ee
Together with \r{logder2} this expresses the correlation function
\r{vcont} in terms of the solution $Y(z)$ of our RHP.

\subsection{Analysis of the RHP}
While the operator-valued RHP defined above determines the correlation
functions completely, its solution appears to be a daunting task in
general. In what follows we concentrate on the leading
term in the asymptotical decomposition of the solution of the RHP in
the region of coupling constant $\frac{\pi}{2}<\g<\frac{2\pi}{3}$. The
reason for this restriction is the following: the upper bound on $\g$
stems from the construction of the ground state of our lattice
regularization. The lower bound ensures that the parameter $p$ defined
in \r{p} will go to infinity in the continuum limit, which essentially
simplifies the analysis of the RHP: it permits us to study the
asymptotical decomposition of the solution of the RHP with respect to
$p$ (recall that $p$ contains the continuum distance as well).Due to
the fact that this parameter will be not only large but diverge the
number of terms in the asymptotical decomposition will be very
small -- in fact we expect only three contributions (see also below).
As we shall show in our analysis of the leading contribution, the
special form of the conjugation matrix in addition to our interest in
partial traces of $Y$ only allows to reduce the RHP to a tractable
scalar one (still containing the auxiliary dual fields, of course). 
The analysis of the subleading terms is technically much more involved
and is currently under investigation. We will report on this work
elsewhere. 

Let us now turn to the calculation of the leading term. First, we note
that the conjugation matrix can be decomposed into the product of an
upper and lower triagonal matrix as follows 
\begin{equation}
\left[G(z|s,t)\right]_{ab} = \sum_{c=1}^4\int_0^\infty ds^\prime
\left[T_1(z|s,s^\prime)\right]_{ac}\left[T_2(z|s^\prime,t)\right]_{cb}\ .
\end{equation}
Here
\begin{equation}
T_1(z|s,t)=\left(\matrix{
1&\a_1(z|s,t)&\a_2(z|s,t)\exp(ipk(z))&\a_3(z|s,t)\exp(ipk(z))\cr
0&1&\a_4(z|s,t)\exp(ipk(z))&\a_5(z|s,t)\exp(ipk(z))\cr
0&0&1&\a_6(z|s,t)\cr
0&0&0&1\cr}\right),
\end{equation}
with matrix elements
\bea
\a_1(z|s,t)&=&\frac{\k\exp(\varphi_4(z))}{1+\k^2\exp(\varphi_4(z)-\varphi_3(z))}
|2,z,s\rangle\langle 1,z,t|\ ,\nn
\a_2(z|s,t)&=&-\frac{1}{\k}\exp(-\varphi_2(z)+\varphi_3(z))
|2,z,s\rangle\langle 2,z,t|\ ,\nn
\a_3(z|s,t)&=&\frac{\k^2\exp(\varphi_1(z)-\varphi_3(z)+\varphi_4(z))}
{1+\k^2\exp(\varphi_4(z)-\varphi_3(z))}|2,z,s\rangle\langle 1,z,t|\ ,\nn
\a_4(z|s,t)&=&-\frac{1}{\k^2}\exp(-\varphi_2(z)+\varphi_3(z)-\varphi_4(z))
|1,z,s\rangle\langle 2,z,t|\ ,\nn
\a_5(z|s,t)&=&\frac{\k\exp(\varphi_1(z)-\varphi_3(z))}
{1+\k^2\exp(\varphi_4(z)-\varphi_3(z))}|1,z,s\rangle\langle 1,z,t|\ ,\nn
\a_6(z|s,t)&=&\frac{\k\exp(\varphi_2(z)+\varphi_1(z)-\varphi_3(z))}
{1+\k^2\exp(\varphi_4(z)-\varphi_3(z))}|2,z,s\rangle\langle 1,z,t|\ .
\eea
Similarly, we find 
\be
T_2(z|s,t)=\left(\matrix{
c_1(z|s,t)&0&0&0\cr
\b_1(z|s,t)&c_2(z|s,t)&0&0\cr
\b_2(z|s,t)\exp(-ipk(z))&\b_4(z|s,t)\exp(-ipk(z))&c_3(z|s,t)&0\cr
\b_3(z|s,t)\exp(-ipk(z))&\b_5(z|s,t)\exp(-ipk(z))&\b_6(z|s,t)
&c_4(z|s,t)\cr}\right).
\ee
The matrix elements of $T_2$ are given by
\bea
\b_1(z|s,t)&=&-\frac{1}{\k}\exp(-\varphi_4(z))|1,z,s\rangle\langle
2,z,t|\ ,\nn 
\b_2(z|s,t)&=&-\frac{\k\exp(\varphi_2(z)-\varphi_3(z))}
{1+\k^2\exp(\varphi_4(z)-\varphi_3(z))}|2,z,s\rangle\langle 2,z,t|\ ,\nn
\b_3(z|s,t)&=&-\k^2\exp(-\varphi_1(z)-\varphi_3(z)+\varphi_4(z))|1,z,s\rangle\langle
2,z,t|\ ,\nn 
\b_4(z|s,t)&=&\frac{\exp(\varphi_2(z))}{1+\k^2\exp(\varphi_4(z)-\varphi_3(z))}
|2,z,s\rangle\langle 1,z,t|\ ,\nn
\b_5(z|s,t)&=&\k\exp(-\varphi_1(z)+\varphi_4(z))|1,z,s\rangle\langle 1,z,t|\
,\nn 
\b_6(z|s,t)&=&-\k\exp(-\varphi_1(z)-\varphi_2(z)+\varphi_4(z))|1,z,s\rangle\langle
2,z,t|\ ,
\eea
and finally
\bea
c_1(z|s,t)&=&1-\frac{\k^2 \exp(\varphi_4(z)-\phi_3(z))}
{1+\k^2\exp(\varphi_4(z)-\varphi_3(z))}|2\rangle\langle 2|\ ,\nn
c_2(z|s,t)&=&1+\frac{1}{\k^2} \exp(-\varphi_4(z)+\varphi_3(z))
|1\rangle\langle 1|\ ,\nn
c_3(z|s,t)&=&1-\frac{1}{1+\k^2\exp(\varphi_4(z)-\varphi_3(z))}|2\rangle\langle
2|\ ,\nn 
c_4(z|s,t)&=&1+{\k^2} \exp(\varphi_4(z)-\varphi_3(z))|1\rangle\langle 1|\ .
\eea

Let us now go through a ``deformation'' of the RHP like for the case
of the Bose gas \cite{ikw:95}.
We define an integral-operator valued function $\wti{Y}(z)$ in the
following way:
\begin{itemize}
\item{} $\wti{Y}(z) = Y(z)$ outside the ``bubble'' defined in Fig 2. In
   particular $\wti{Y}(z) = Y(z)$ for $z\rightarrow 0,\infty$, which will
   be important later.
\item{} $\wti{Y}(z)=Y(z)T_1(z)$ in the region enclosed by the real axis and
   the contour $\Gamma_1$. Note that in this region ${\rm Im}k(z)\geq 0\
   \forall z$.
\item{} $\wti{Y}(z)=Y(z)[T_2(z)]^{-1}$ in the region enclosed by the real
   axis and the contour $\Gamma_2$. Note that in this region ${\rm
   Im}k(z)\leq 0\ \forall z$.
\end{itemize}
It can be easily seen that the function $Y(z)$ defined in the above
way has the following properties:
$\wti{Y}(z)$ is analytic in the whole complex plane with the exception of
the contours $\Gamma_1$ and $\Gamma_2$. On the contours $\Gamma_j$
$\wti{Y}$ satisfies the conjugation equations 
\bea
(\wti{Y})^-(z)&=&\wti{Y}^+(z) T_1(z)\ ,\qquad z\in\Gamma_1\ ,\nn
(\wti{Y})^-(z)&=&\wti{Y}^+(z) T_2(z)\ ,\qquad z\in\Gamma_2\ .  
\eea
Since we are only interested in the asymptotic behaviour of the determinant
for $p\gg1$ we can use the fact that in this limit $T_{1(2)}$ become
blockdiagonal in the vicinity of the contour $\Gamma_{1(2)}$ from which we
find that
\be
Y(z)\sim \left(\matrix{\widetilde{\Phi}_1(z)&0\cr 0 &
\widetilde{\Phi}_2(z)\cr}\right)\ .
\ee
Here $\widetilde{\Phi}_j(z)$ are solutions to $2\times 2$ operator-valued
RHPs
\be
\widetilde{\Phi}_j^-(z)=\widetilde{\Phi}_j^+(z)*G_j(z)\ ,\quad j=1,2
\ee
with the same conjugation contour $C$ as the original RHP and
conjugation matrices 
\be 
G_1(z)=
\left(\matrix{1-|2\rangle\langle 2|& 0\cr
        0& 1-|1\rangle \langle 1|\cr}\right)
+\left(\matrix{0&\frac{\exp(\varphi_3(z))}{\kappa}|2\rangle\langle 1|\cr
-\frac{\exp(-\varphi_4(z))}{\kappa}|1\rangle\langle 2|&
(1+\frac{\exp(\varphi_3(z)-\varphi_4(z))}{\kappa^2})|1\rangle\langle
1|\cr}\right)\ ,
\label{g1}
\ee

\bea
G_2(z)&=&
\left(\matrix{1-|2\rangle\langle 2|& 0\cr
        0& 1-|1\rangle \langle 1|\cr}\right)\nn
&&+\left(\matrix{0&
{\kappa\exp(\varphi_1(z)+\varphi_2(z)-\varphi_3(z))}|2\rangle\langle 1|\cr
-{\kappa\exp(-\varphi_1(z)-\varphi_2(z)+\varphi_4(z))}|1\rangle\langle 2|&
{(1+\kappa^2\exp(\varphi_4(z)-\varphi_3(z)))}|1\rangle\langle 1|\cr}\right).
\nonumber
\label{g2}
\eea

Using the fact that $G_j(z)$ form representations of $GL(2|C)$ we can
now calculate the determinants of $G_j(z)$ as is shown in the appendix
\be
\det(G_1(z))= \exp(-\a+\varphi_3(z)-\varphi_4(z))\ ,\qquad
\det(G_2(z))= \exp(\a-\varphi_3(z)+\varphi_4(z))\ .
\ee
The scalar RHPs for the determinants 
\[
   \det(\widetilde{\Phi}_j^-(z))
	=\det(\widetilde{\Phi}_j^+(z)) \det(G_j(z))\ ,\quad j=1,2
\]
is now easily integrated to give
\bea
\det(\widetilde{\Phi}_1(z))&=&\exp(-\frac{1}{2\pi i}\int_\A^\B dz_1\
\frac{-\a+\varphi_3(z_1)-\varphi_4(z_1)}{z_1-z})\ ,\nn
\det(\widetilde{\Phi}_2(z))&=&\exp(-\frac{1}{2\pi i}\int_\A^\B dz_1\
\frac{\a-\varphi_3(z_1)+\varphi_4(z_1)}{z_1-z})\ .
\label{scalar}
\eea

\section{Leading Term in the Asymptotics of the Correlator}
\label{sec:asy}
Let us now relate the solution of the scalar RHPs to the logarithmic
derivative of $\det(1+\frac{1}{2\pi}\widehat{V}^T)$. The contribution due
to $C^{(0)}$ in \r{logder2} can be obtained from \r{contr2} and \r{scalar}
as
\bea
\frac{1}{2}\sum_{k=3}^4\int_0^\infty ds C^{(0)}_{kk}(s,s)&=&
\lim_{z\to\infty}\frac{iz}{2}\ln\left(\det(\widetilde{\Phi}_2(z))\right)\nn
&=&\frac{1}{4\pi}\int_\A^\B dz[\a-\varphi_3(z)+\varphi_4(z)]\ .
\label{ccontr1}
\eea
Similarly, the second contribution in \r{logder2} is with \r{contr2}
\bea
  &&\frac{1}{2}\sum_{k=3}^4\int_0^\infty ds
	\left([I-iB^{(1)}]*C^{(2)}\right)_{kk}(s,s) =
	-\frac{i}{2}\frac{d}{dz}\bigg|_{z=0}
	\ln[\det(\widetilde{\Phi}_2(z))]\nn
  &&\qquad= \frac{1}{4\pi}\int_\A^\B dz\ 
	\frac{\a-\varphi_3(z)+\varphi_4(z)}{z^2}\ .
\label{ccontr2}
\eea
Combining these to the leading asymptotical bahaviour of
$\partial_p\ln(\det(1+\frac{1}{2\pi}\widehat{V}^T))$ and using the fact
that they are $p$-independent we obtain \bea
\det(1+\frac{1}{2\pi}\widehat{V}^T)&=& A\exp(\frac{\a
p\sinh(2\La)}{\pi})\exp(\frac{p}{4\pi}\int_\A^\B dz\ (1+\frac{1}{z^2})
(\varphi_4(z)-\varphi_3(z)))\nn &=&A\exp(\frac{\a p\sinh(2\La)}{\pi})
\exp(\frac{p}{\pi}\int_{-\La}^\La d\l \cosh(2\l)
(\varphi_4(\l)-\varphi_3(\l)))\ , \eea
where $A$ is a $p$-independent constant and where in the last step we have
changed back to the original $\l$-variables.  Decomposing the combination
of dual fields into ``momenta'' and ``coordinates'' and using the
commutation relations \r{dfcom} we find
\be
\varphi_4(\l)-\varphi_3(\l)=P(\l)+Q(\l)\ ,\qquad [Q(\mu),P(\l)]=0\ .
\ee
This enables us to trivially evaluate the expectation value with respect to
the dual fields in this approximation: the dual fields are found not to
contribute at all leading to the following result for the leading
asymptotical behaviour of the correlator
\be
\langle\Omega|\exp(\alpha Q_1(n))|\Omega\rangle \sim
\widetilde{A} \exp(\frac{\a p}{\pi}\sinh(2\La))\ ,
\label{finalres}
\ee
where $\widetilde{A}$ is a constant independent on $p$.  

We will now argue that the approximation \r{finalres} is too crude due to
the fact that we have neglected the influence of the dual fields in the
{\sl subleading} factors in the solution of the RHP. We expect
the final answer for the solution of the RHP to be of the form
\bea
\det(1+\frac{1}{2\pi}\widehat{V}^T)&=&
C(\{\varphi_j\})\ \exp(\zeta(\{\varphi_j\})\ln(p))\ \nn
&&\times \exp\left(\frac{\a p\sinh(2\La)}{\pi}\right)
\exp\left(\frac{p}{\pi}\int_{-\La}^\La d\l \cosh(2\l)
(\varphi_4(\l)-\varphi_3(\l))\right)\ ,
\label{conjec}
\eea 
where we keep in mind that $p\rightarrow\infty$ as the lattice spacing
$\Delta\rightarrow 0$. In \r{conjec} $C$ is $p$-independent and we have
conjectured that the subleading term in the solution of the RHP is a
power-law in $p$. Evaluating the expectation value of \r{conjec} in
the dual bosonic Fock space {\sl the dual fields will contribute in the
exponential term}, {\sl i.e}
\begin{eqnarray}
   \langle\exp(\a Q_1(n))\rangle &\sim&
   \nddv\det(1+\frac{1}{2\pi}\widehat{V}^T)\dv \nn
   &=& \widetilde{C}\ p^{\widetilde{\zeta}}
	\exp\left(\widetilde{m}p\right)
	\exp\left(\xi p\ln(p)\right)
	\exp\left([\frac{\sinh(2\La)}{\pi}+\omega]\a p\right).
\label{answer}
\end{eqnarray}
Here $\widetilde{\zeta}$, $\omega$ and $\widetilde{m}$ are functions of
$\gamma$, the soliton mass {\sl etc}. For this answer to be of the correct
qualitative form, the following conditions have to be satisfied:
\begin{itemize}
\item $\xi=0$, as the leading asymptotic behaviour should be $\exp({\rm
      const}\ p)$.
\item
 The last factor in \r{answer} has to be cancelled by a suitable
 regularization procedure for the result to make sense. In the continuum
 limit we have \r{Qcont} which implies that
\begin{equation}
   \langle\exp(\a Q_1(n))\rangle \rightarrow 
	\langle\exp\left(\frac{2\a}{\b}[u(x)-u(0)]\right)\rangle\ 
	\times\ 
	\exp\left({\a x\over\Delta}\ {\pi-\gamma\over2\gamma}\right).
\label{contlim}
\end{equation}
  We see that this expression contains a divergent factor depending both
on $\a$ and on the distance $x$. We now adjust our ``cutoff'' $\La$ in such
a way that the divergent factor in \r{answer} precisely reproduces the
divergent factor in \r{contlim}, {\sl i.e.}
\[
   \exp\left(\frac{\a x}{\Delta}\frac{\pi-\g}{2\g}\right)=
   \exp\left([\frac{\sinh(2\La)}{\pi}+\omega]\a p\right).
\]
If $\omega=0$ this leads to the following relation between the ``cutoff''
$\La$ and the lattice spacing $\Delta$
\begin{equation}
   \exp(2\La) =\Delta^{-2\frac{\pi-\g}{\pi}}
   \left(\frac{8\pi(\pi-\g)}{c^2\sin(\g)\g}\right)
\label{ladel}
\end{equation}
with a finite constant $c$. The procedure outlined above fixes the relation
between $\La$ and $\Delta$ and thus between the divergent part of $p$ and
$\La$. Note, however, that the result \r{ladel} for this relation is not
consistent with the requirement $S\cosh(2\l)\ll 1$, which we have used in
order to simplify the kernel of $\widehat{V}$ in section \ref{sec:cont}.
Therefore the assumption $\omega=0$ has to be wrong and we do need a
$\Lambda$-dependence of $\omega$ instead which corrects \r{ladel}.
\end{itemize}

\section{Summary and Conclusion}

In this paper we have applied the method of \cite{vladb} to
correlation fucntions in the \sG\ model. In order to deal with the UV
divergences we used an integrable lattice regularization of the \sG\
model to derive a determinant representation for quantum correlation
functions. We then took the continuum limit and obtained a determinant
representation for the \sG QFT. Furthermore we embedded the
determinant in a system of integrable integro-differential equations
which we showed to be associated with an operator-valued
Riemann-Hilbert problem. The quantum correlation function was
expressed in terms of the solution of this RHP. We then presented a
general approach to obtain the leading asymptotical behaviour of the
solution of the RHP, which in turn yields the leading term in the
asymptotics of the quantum correlation function. We showed that the
subleading terms in the asymptotical decomposition are essential for
obtaining explicit expressions for the asymptotics of the correlation
function due to the presence of the dual quantum fields. For the case
at hand there appear to be only two subleading terms in the
asymptotical decomposition which is very encouraging!
The analysis of the subleading terms is a difficult mathematical
problem by itself and we will report on it in a separate publication.

\section*{Acknowledgements}
The authors are grateful to A.\,G.\,Izergin and N.\,M.\,Bogoliubov for
useful discussions. This work was partially supported
by the Deutsche For\-schungs\-gemein\-schaft under Grant No.\ Fr~737/2--1
and by the National Science Foundation (NSF) under Grants No.\
PHY-9321165 and No.\ DMS-9501559.
F.H.L.E.\ is supported by the EU under Human Capital and Mobility
fellowship grant ERBCHBGCT940709.

\appendix
\section{$Gl(2|C)$ Representation by Integral Operators}

In this appendix we show how to essentially simplify the analysis of
the operator-valued RHP through the use of $GL(2|C)$ representation
theory. We closely follow the discussion of \cite{ikw:95}.

Let us consider an integral-operator valued $2\times 2$ matrix with kernel

\begin{equation}
\Op(s,t)=\left(\begin{array}{cc}
			\Op_{11}(s,t)&\Op_{12}(s,t)\\&\\
			\Op_{21}(s,t)&\Op_{22}(s,t)
			\end{array}\right), \; s,t\in[0,\infty).
\end{equation}
Multiplication of integral-operator valued matrices $\Op$ and ${\cal
P}$ is defined in the usual way as
\be
\left[{\cal OP}\right]_{ij}(s,t)=\sum_{k=1}^2
\int_{0}^{\infty}dr\ \Op_{ik}(s,r){\cal P}_{kj}(r,t)\ ,\quad
i,j=1,2.
\ee
The left (right) action of the integral operators $\Op_{ij}$ on fuctions
defined on the interval $[0,\infty)$ is given by
\be
\Op_{ij}*f\bigg|_s= \int_0^\infty dt\ \Op_{ij}(s,t)\ f(t)\ ,\quad
g*\Op_{ij}\bigg|_t= \int_0^\infty ds\ f(s)\ \Op_{ij}(s,t)\ .
\ee
Let us now construct a special class of such operators
$\widehat{\Op}$ which form a representation of  $Gl(2,{\bf C})$:
we start with two pairs of functions $(\alpha(s),\ \beta(s))$ and
$(A(s),\ B(s))$ on $[0,\infty)$ which we represent in Dirac notation as
$\alpha(s)\equiv|1\rangle$, $\beta(s)\equiv|2\rangle$,
$A(s)\equiv\langle 1|$ and $B(s)\equiv\langle 2|$. These functions are
chosen in such a way that
\begin{equation}
\langle1|1\rangle\equiv\int_{0}^{\infty}dsA(s)\alpha(s)=1=
\langle2|2\rangle\equiv\int_{0}^{\infty}dsB(s)\beta(s).
\end{equation}
In this notation we may write left multiplication by $\widehat{\Op}_{ik}$ as
\begin{equation}
\widehat{\Op}_{ik}|1\rangle=\int_{0}^{\infty}dt\Op_{ik}(s,t)\alpha(t).
\end{equation}
Observe now that one may define a representation
 $\widehat{\cal{A}}$ of $Gl(2,{\bf C})$ in terms of integral operators
via
\begin{equation}
M\!\in Gl(2,{\bf C})\longmapsto{\widehat{\cal A}}(M)=\left(\begin{array}{cc}
					 I-|1\rangle\langle1| & 0\\&\\
					 0 & I-|2\rangle\langle2|
                                         \end{array}\right)
				+	 \left(\begin{array}{rr}
	 M_{11}|1\rangle\langle1| & M_{12}|1\rangle\langle2|\\&\\
			 M_{21}|2\rangle\langle1| & M_{22}|2\rangle\langle2|
                                         \end{array}\right).
\end{equation} 
Here $M_{11}$, $M_{12}$, $M_{21}$ and $M_{22}$ are complex numbers and
$I$ is the identity operator in the space of integral oprators on
$[0,\infty)$.
Multiplication by the integral operators $|1\rangle\langle1|$,
$|1\rangle\langle2|$, $|2\rangle\langle1|$ and $|2\rangle\langle2|$
is given by {\sl e.g}
\begin{equation}
|1\rangle\langle2|f(s)=\left(\int_{0}^{\infty}dsB(s)f(s)\right)|1\rangle.
\end{equation}
Therefore $|i\rangle\langle j|$ act like projectors on the ``states''
$|i\rangle$ and $\langle j|$.

In particular identities like $\left[ I-|1\rangle\langle1|\right]
|1\rangle\langle1|=0$ are seen to hold.
Indeed for any $M$, $N$ $\in$ $Gl(2,{\bf C})$ the representation
$\widehat{\cal A}$ has the following properties
\bea
{\rm (P1)}\quad&&\widehat{\cal A}(MN)=\widehat{\cal A}(M)\widehat{\cal A}(N)\ ;
\ \widehat{\cal A}(I)=I \ ;\ \widehat{\cal A}(M^{-1})=
\widehat{\cal A}^{-1}(M)\nonumber\\[.5cm]
{\rm (P2)}\quad&&{\rm Tr}\left(\widehat{\cal A}(M)-\left(\begin{array}{cc}
					 I-|1\rangle\langle1| & 0\\&\\
					 0 & I-|2\rangle\langle2|
        \end{array}\right)\right)=\tr M=M_{11}+M_{22}\ ,\nonumber\\[.5cm]
{\rm (P3)}\quad&&{\rm Det}\widehat{\cal A}(M)=\det M=
M_{11}M_{22}-M_{12}M_{21}\label{detA}\ .
\label{rep}
\eea
Properties (P1) and (P2) can be established by direct computation
using the rules given above.
Property (P3) shows that the determinant of the integral
operator ${\cal A}$ is simply equal to the determinant of the
$2\times2$ matrix $M$, which is quite remarkable. It is established by
expressing the determinant as a trace {\sl via}
$\ln{\rm Det} {\cal A}= {\rm Tr}\ln {\cal A}$, then using (P1) in the
expansion of the logarithm, using (P2) to express the operator trace in
terms of the matrix trace, and finally expressing the sum over traces
back as determinant of the matrix $M$.

It can be easily checked that the representation \r{g1} of the
conjugation matrices $G_j(z)$ is precisely of the above form (here $z$
plays the role of a parameter), which in turn allows us to evaluate
the determinants of the conjugation matrices.

\newpage

\section*{Figures}
\begin{figure}[h]
\begin{center}
\leavevmode
\epsfbox{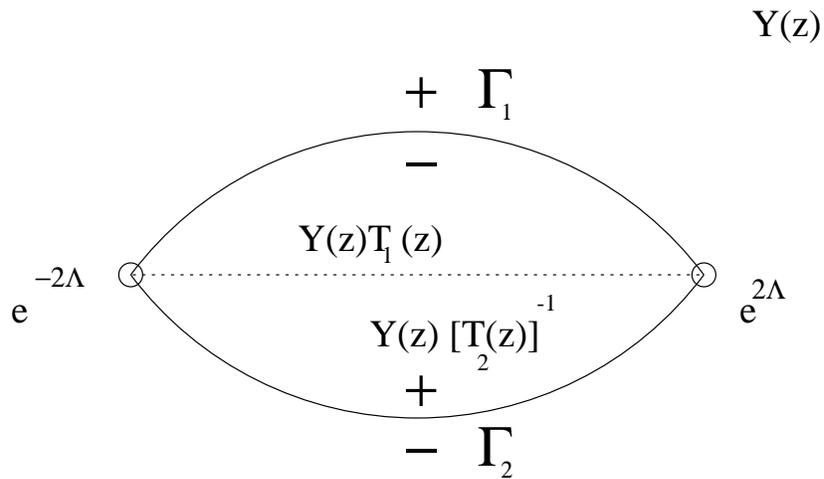}
\end{center}
\caption{Deformation of the Conjugation Contour for the RHP}
\end{figure}

\newpage
\setlength{\baselineskip}{13pt}

\end{document}